Laboratoire de Physique Nucléaire et de Hautes Énergies

CNRS - IN2P3 - Universités Paris VI et VII

# Analysis of Purity Probes
# H1 Liquid Argon Calorimeter

E. Barrelet, T.P. Yiou

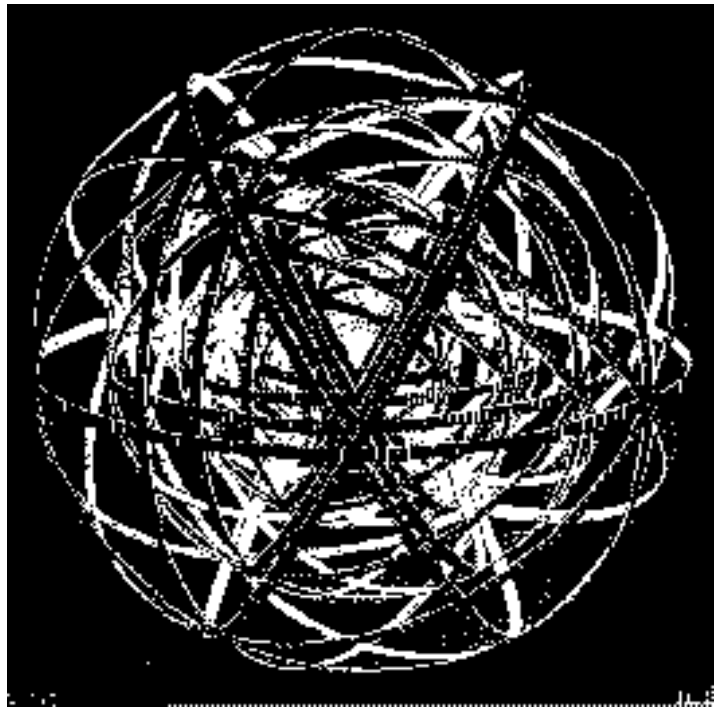

4, Place Jussieu - Tour 33 - Rez-de-Chaussée
75252 Paris Cedex 05

Tél : 33(1) 44 27 63 13 - FAX : 33(1) 44 27 46 38

Abstract:
The sensitivity of our liquid argon purity measurement -around 0.03%- leads us to refine the tools used classically in this field. First we introduce an analytical form describing the spectrum of a $^{207}$Bi source for different values of purity and ionisation chamber gap. Second we analyse a surprising new effect: the variation of the ionisation yield of this source with the liquid argon temperature. Third we use our data to refine the charge collection models which relate electron recombination and attachment cross-sections to the electrical field.

**Introduction**

During 8 years, we have tracked the yield of $^{207}$Bi and $^{241}$Am radioactive sources with 9 ionization chambers disseminated inside the H1 liquid argon (LAr) calorimeter. This was a group's work as acknowledged later. Our goal was to monitor the purity of LAr with a relative precision better than 0.1% in order to detect a pollution before it might spoil the 1% accuracy of the H1 calorimeter. The description of this system and its use by H1 is the subject of a paper to be submitted to NIM. This report details the more recent tools that we had to build for this analysis.

Our first section will study the spectra delivered by our purity probes. They will be successfully represented by an *analytical function* ($S_{Bi}(q)$), which justifies the simple estimators used to reduce by a huge factor the amount of data taken. It will also allow us to quantify the systematic errors, below 1%, affecting our purity estimates. Moreover it will provide us with quantitative tools for our third section.

Our second section will gather all accessible information concerning an unforeseen "*liquid argon temperature effect*" (LAr-T). The origin of this effect remains mysterious, but it is well described as a variation of the ionization yield of $\approx 1.5\%/°K$ affecting $^{207}$Bi sources. It is not directly related to the temperature dependence of either electron mobility or LAr density. It could affect equally LAr electromagnetic calorimeters, but this has yet to be proven.

Our last section is devoted to the *"high voltage curve"* (HV-curve) method using the measurement of charge collection efficiency as a function of field strength *E* in order to infer the W-value and the electron mean free path. Our data, at least 10 times more precise than earlier data, obliged us to improve current models particularly in the asymptotic region $E \to \infty$. We had to replace the classical electron attachment cross-section proportional to $E^{-1}$ by a function showing saturation for $E \to \infty$. We refute also earlier publications which unfold explicitly the recombination and the attachment HV-curves, because:

1- a recombination parameter is not independent of the type of source used as they assume.
2- unfolding is a mathematically ambiguous procedure within a reasonable range of *E*-field and experimental accuracy.

As a result of this analysis the purity measurements at medium and at high electrical field are reconciled. This lead us to represent HV-curves using an "*impurity concentration*" variable based on the real electron attachment cross-section, which is valid for all values of *E*.

# 1  Analysis of $^{241}$Am and $^{207}$Bi spectra

The goal of the analysis is to measure the *electron mean free path* $\lambda_e$ due to impurities in the liquid argon.
The strategies for analyzing $^{241}$Am and $^{207}$Bi ionization spectra are different:

- The Americium emits mainly 5.49 MeV α-rays. Its ionization spectrum, presenting a single isolated peak, is parametrized by a mean charge $Q_\alpha$ and an effective width $\sigma_\alpha$ (R.M.S.). (We compute equally an edge estimator, which is more precise but depends both on $Q_\alpha$ and $\sigma_\alpha$).
  We operate at 20 KV/cm in order to minimize recombination losses (75%). Ignoring possible fluctuations of the recombination, $Q_\alpha$ monitors the **relative variation of** $\lambda_e$.

- The Bismuth e- and γ-ray spectrum, more complex, is resumed by two peak and one edge estimators represented in figure 1. Its main advantage is to be directly **comparable to H1** electromagnetic calorimeter data (same range of specific ionisation and field strength) and to give access to the **absolute value of** $\lambda_e$ by fitting the relative variation of $\lambda_e$ with field strength (cf. 3.1).

## 1.1  The ionization spectrum of $^{207}$Bi

We have shown in figure 1 eight spectra taken at 2 days interval during the test of an H1 calorimeter stack in a CERN beam from April 25th to May 9th 1990. The concentration of impurities was rising very rapidly, allowing us to calibrate our device on a range of $\lambda_e$ larger than seen afterwards in H1 (from year 1991 to 1999). Moreover the response of our probes could be cross checked with the response of the H1 calorimeter stack sitting in the same cryostat during the same period.
The 2 prominent peaks $\gamma_1$ and $\gamma_2$ are due to the electron conversion of the 2 lower energy gammas emitted by $^{207}$Bi ($E_{\gamma_1} = 0.569 MeV$  $E_{\gamma_2} = 1.063 MeV$  $E_{\gamma_3} = 1.77 MeV$). The splitting of the $\gamma_1$ peak into K(482 keV) and L(554 keV) electron lines is marked by a shoulder on the $\gamma_1$ falling edge. No splitting is seen on the $\gamma_2$ peak, which is wide because the range of corresponding electrons is large compared to the gap. The continuous background is due to the Compton scattering of all 3 gammas.
Looking at our sequence of 8 spectra, both γ peaks are shifting leftward with time because $\lambda_e$ is decreasing with rising impurity level, while the electronic calibration peak stays constant.

## 1.2  $^{207}$Bi spectral analysis

We have fitted the $^{207}$Bi energy spectrum using an analytical function $S_{Bi}(q)$ given in 5.2, which reproduces explicitly the variations of the $^{207}$Bi spectrum with varying gap size and impurity concentration. The free parameters of our model have been determined on the 25/04/90 spectrum in figure 1 by eyeball fit. The LAr gap *D* being fixed at 6mm, the best value for the electron mean free path was $\lambda_e$ = 4.32 mm. The rate of the $\gamma_2$ electron conversion peak being 773 hz, we fixed the ratio of L to K electrons in this peak to be 25%, following an independent calibration of our source done with a solid state counter[1]. We found the $\gamma_1^K$ and $\gamma_1^L$ rates to be respectively 17.2% and 7.0% of $\gamma_2^K$. These numbers looks compatible with the 21% and 6% obtained by the calibration, since 2$^{nd}$ order effects are not the same in different ionization media (back-scattering etc....). From the width of the $\gamma_2$ peak, we can deduce a "practical range" $R(E) = 4.5 \pm 0.2$ mm ↔ 6.3 g/cm² for 1 MeV electrons in argon. This compares to a range of 5.5 g/cm² in the PDG tables. The Compton events for $\gamma_1$ and $\gamma_2$ are seen respectively with integrated rates 3.7 and 1.4 times higher than the corresponding electron conversion rates and the $\gamma_3$ Compton's represents 12% of the total. These rates determine the coefficients $A_i$ in 5.2 Equation 18. Experimental determinations of the Compton edges $E_1^{edge}$ and $E_2^{edge}$ are compatible with their kinematical values at a 1% and 5% level. Finally, as seen on figure 2, just by changing $\lambda_e$ from 4.32 mm into 3.14 mm, one changes exactly the first spectra of figure 1 into the last.

---

[1]  thanks to G.Roubaud, Radioprotection group, CERN

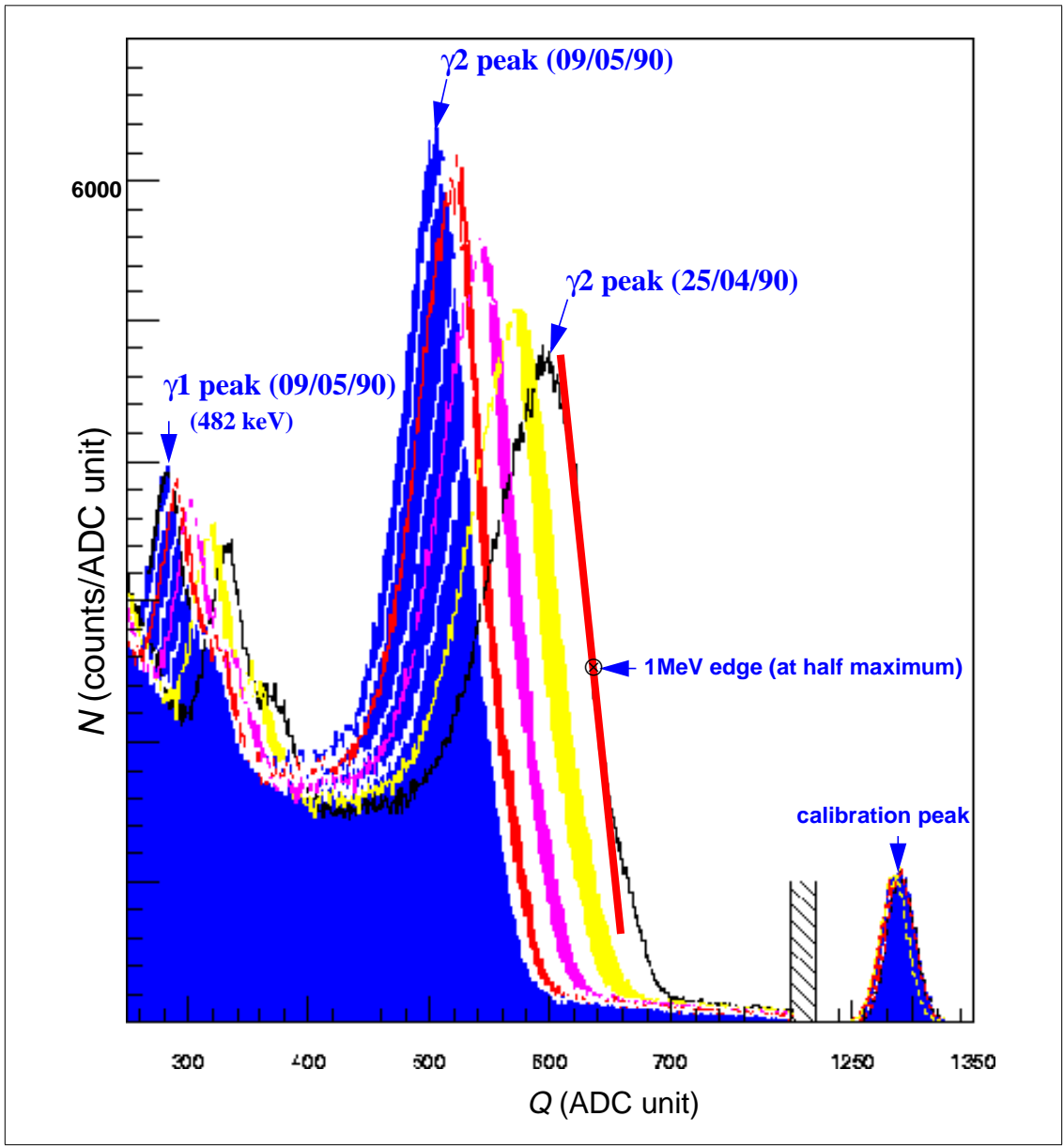

**Figure 1:** Eight 207Bi spectra accumulated during a single run at CERN (april-may 90)

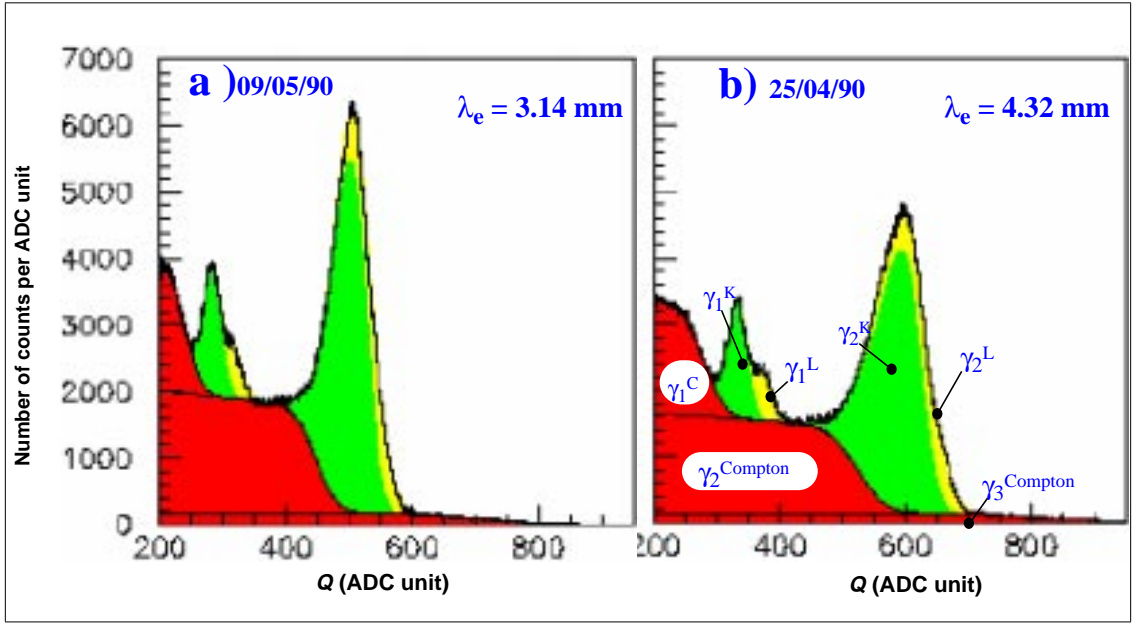

**Figure 2:** Comparison of the analytical function $S_{Bi}(Q)$ with spectra of a $^{207}$Bi probe with 6 mm gap. a) data of 09/05/90 b) data of 25/04/90. Areas corresponding to the three electron conversion mechanisms (K, L, Compton) are filled with different shades

### 1.3 purity estimators

The most precise ionization charge measurement is done by fitting the falling edge of $\gamma_2$ with a straight line around 1 MeV, where the Compton background is minimum. On this line, we determine either the point having a given ordinate or corresponding to a constant fraction of the peak height. The fraction 1/2, used later, is materialized in figure 1. Statistical accuracy reached by this "1 MeV edge" estimator within a 5 minutes data taking is ≈0.03% RMS. An other independent "482 KeV peak" estimator of the charge yield is obtained by fitting a Gaussian peak plus a polynomial background in a window around the $\gamma_1$ peak. In figure,3 we have used the parametrization of our spectra to evaluate the quality of the "1 MeV edge" and "482 KeV peak" estimators and to compare them with the "$\gamma_2$ peak" estimator (parabolic approximation inside $\gamma_2$ FWHM interval). These first 2 estimators follow within 1% the behavior of an ideal point-like deposition on the cathode given by the function $\eta(0, \lambda_e/D)$, introduced in Appendix 1. This approximation deviates below $\lambda_e$ = 2.4 mm because the signal, attenuated by a factor >3, is affected by electronic noise. If we fix the energy scale on the 482 KeV peak, then the 1 MeV edge estimator reads 1.02 MeV. On the contrary the $\gamma_2$ peak is clearly not "point-like" and varies from 0.88 to 0.83 MeV.

### 1.4 conclusion

The remarkable accuracy of the analytical representation of the $^{207}$Bi spectrum and of the electron mean free path estimators justifies the design of a single gap purity probe. It rules out, in our opinion, the use of a double gap "Frisch grid" assembly for this application. (Frisch grid and high purity are essential for liquid argon spectroscopy of electron lines above 500 keV). Single gap means simplicity and higher field strength. Moreover, with Frisch's device, the effect of impurities in the conversion gap and of grid transparency are not controllable[2]. This is a major obstacle for studying the HV-curves with an accuracy comparable to ours (better than 0.1% as seen in Section 3).

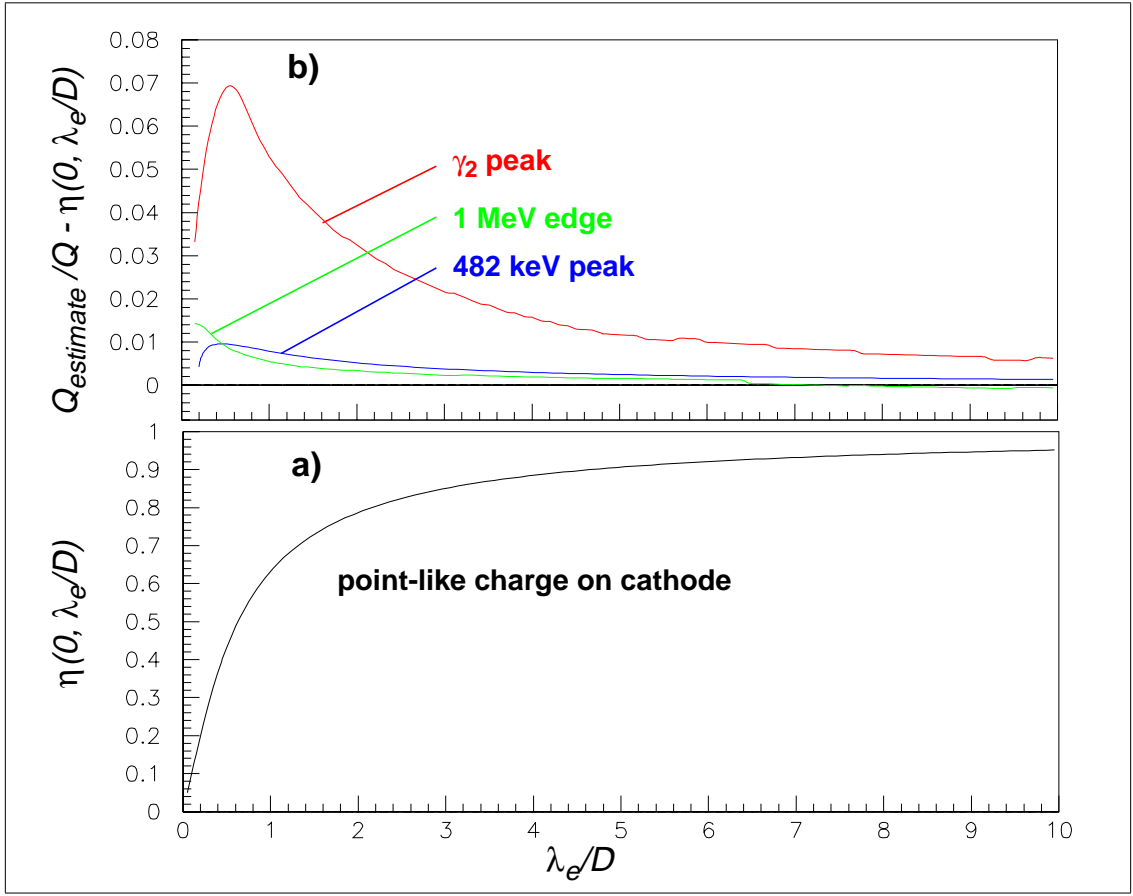

**Figure 3:** a) Charge collection efficiency for a point-like charge $Q$ deposited on the cathode $Q_{point-like}/Q = \eta(0, \lambda_e/D)$. ($\eta$ function is defined in Appendix 1, Equation 11); b) Discrepancy function $Q_{estimator}/Q - \eta(0, \lambda_e/D)$ for 482 keV peak, 1 MeV edge and $\gamma_2$-peak estimators applied to our spectral function ($D$=6mm). [curves based on the analytical representation of $^{207}$Bi spectrum]

## 2 Variation of ionization yield with temperature

This section is devoted to a "*liquid argon temperature effect*" (LAr-T) observed in the course of our measurement. The analysis is based essentially on the data recorded during a shutdown of HERA machine in january 1995. Deliberate modifications of H1 LAr calorimeter temperature were made. This "temperature scan" was sufficient to produce correction coefficients applicable to the analysis of our purity data. However more experiments are needed to understand fully this LAr-T effect. Understandably they cannot be done during the normal operation of the H1 calorimeter.

### 2.1 LAr-T effect on $^{207}$Bi spectrum

On figure 4 we see that the response of the "1 MeV" charge estimator follows closely the evolution of temperature in the LAr cryostat during the temperature scan. Let us note that two estimators determined on different regions of the same spectrum display the same pattern (with larger statistical errors) although they are not sensitive to the same systematic errors. The "1 MeV", "$\gamma_2$ height" and "$\gamma_1$ peak" estimators measure respectively the high energy edge, the height of the central region of the $\gamma_2$ peak (average number of events par ADC unit) and the center of the gaussian + background fit of the 482 KeV line. This observation excludes the possibility of an artifact linked to an unexpected modification of the spectrum (e.g. electronic noise etc....). It shows that the whole spectrum is shifted proportionally to energy.

---

[2] For instance Biller et al.[4] uses drift distances from 1 to 4 cm. In order to test our $E$ and $\lambda_e$ values with their apparatus, one should either apply up to 40 kV or measure charges attenuated by a factor 1.6x10$^{-6}$. In reference[4], spectra vary wildly with drift distance. They are not analysed. This could explain a wrong behaviour of their fits ($\lambda_e(E=0) \neq 0$ and $1/\lambda_e(E \to \infty) = 1+1/E$) and the failure of cross-section scaling.

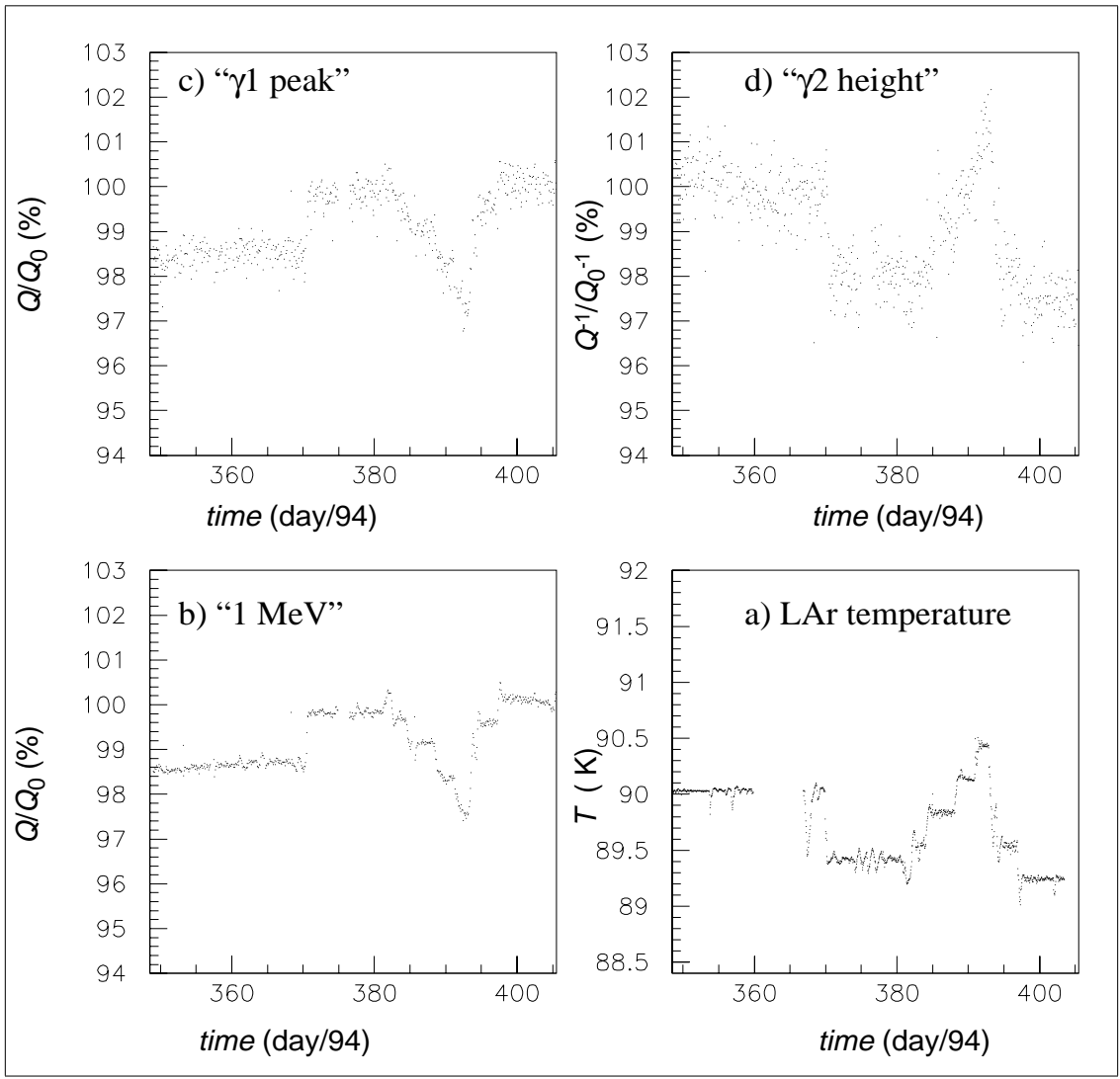

**Figure 4:** a) Variation of liquid argon temperature during temperature scan (Dec. 94 to Feb. 95) b,c,d) Induced variation of three $^{207}$Bi charge estimators for probe 2 (gap $D$= 6mm)

In order to quantify the LAr-T effect we made a regression analysis for each probe, such as the one shown in figure 5 which gives $\Delta Q/Q \Delta T = -1.8\%/°K$. This result is obtained for the 6 mm gap probe. For the 6 other probes with a 4 mm gap, the same analysis yields $\Delta Q/Q \Delta T = -1.28 \pm 0.12\%/°K$.

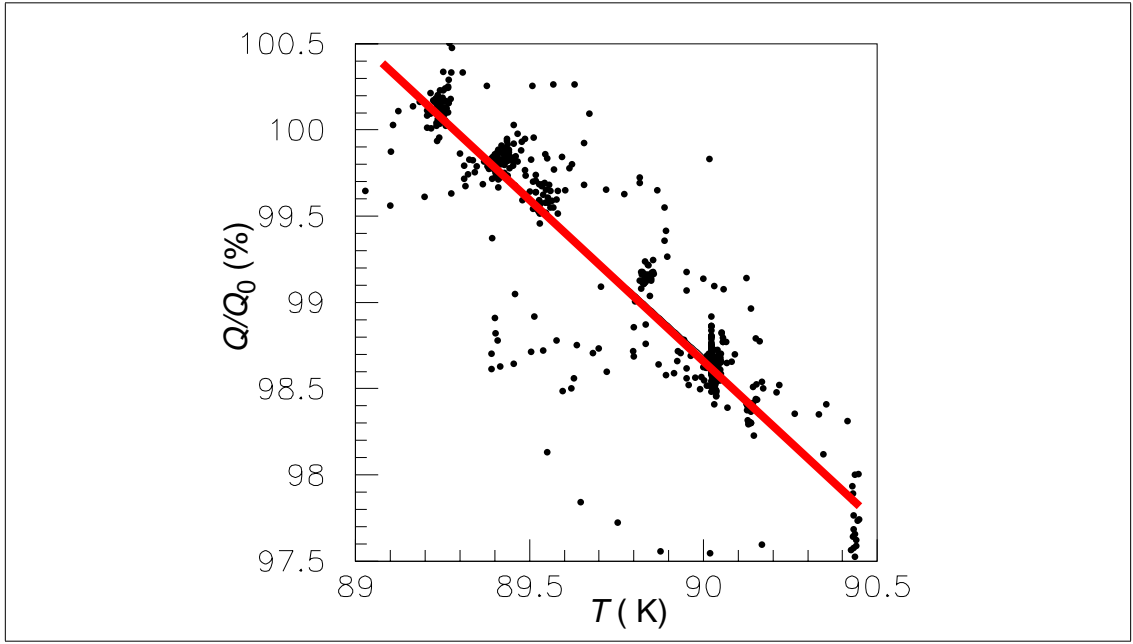

**Figure 5:** $Q/Q_0$ ($^{207}$Bi 1 MeV estimator; 6mm probe) vs. average liquid argon temperature

### 2.2 LAr-T effect on $^{241}$Am spectrum

During the temperature scan of january 1995 no LAR-T effect at all was seen on both $^{241}$Am probes. This is clearly seen in figure 6 by comparing $^{241}$Am to $^{207}$Bi. The observation of such a different behavior of alpha and beta probes eliminates all interpretations based on impurities or electronic and readout artifacts: the same number of electrons are generated by each type of probe, they are attached on the same impurities and they go through the same readout system. ($^{241}$Am probe sensitivity to impurities is a third of that of $^{207}$Bi 6 mm probe).
At this stage we are left with two possibilities of differentiating alpha and beta sources:

- the high density of ionisation of alphas
- the distribution of leftover $Ar^+$ ions across the gap for betas.

### 2.3 Noise introduced by thermal fluctuations

Our most precise charge collection measurement -the 1 MeV estimator- is affected by the thermal fluctuations of H1 cryostat. Under normal conditions (sampling frequency ≈ 1 hour$^{-1}$), noise varies from 0.025 to 0.05 °K(RMS) according to the point of measure as seen in figure 7 b) and with non-gaussian tails reaching 0.5 to 1 °K. These tails are mainly due to the delivery of cold liquid nitrogen from the factory every 2 days or so. The charge vs. temperature correlation is clearly seen in fig. 7 c for these tails.
For the gaussian core of charge and temperature distributions, a strict causal relationship is difficult to prove because the response of each probe and each thermometer to the thermal fluctuations of the cryostat are different, but they show the same frequency spectrum.
The $\sigma_\eta = 0.076\%$(RMS) charge fluctuation (gaussian fit in figure 7a) can be compared to the corresponding thermal noise $\sigma_T = 0.04$ °K. Actually $\sigma_\eta/\sigma_T$ is approximately equal to the coefficient $\kappa = \Delta Q/Q\Delta T = -1.8\%/°K$ determined in section [2.1]). Moreover the long term variation of $\sigma_\eta$ is well represented by the quadratic average of $\kappa.\sigma_T$ and the statistical error

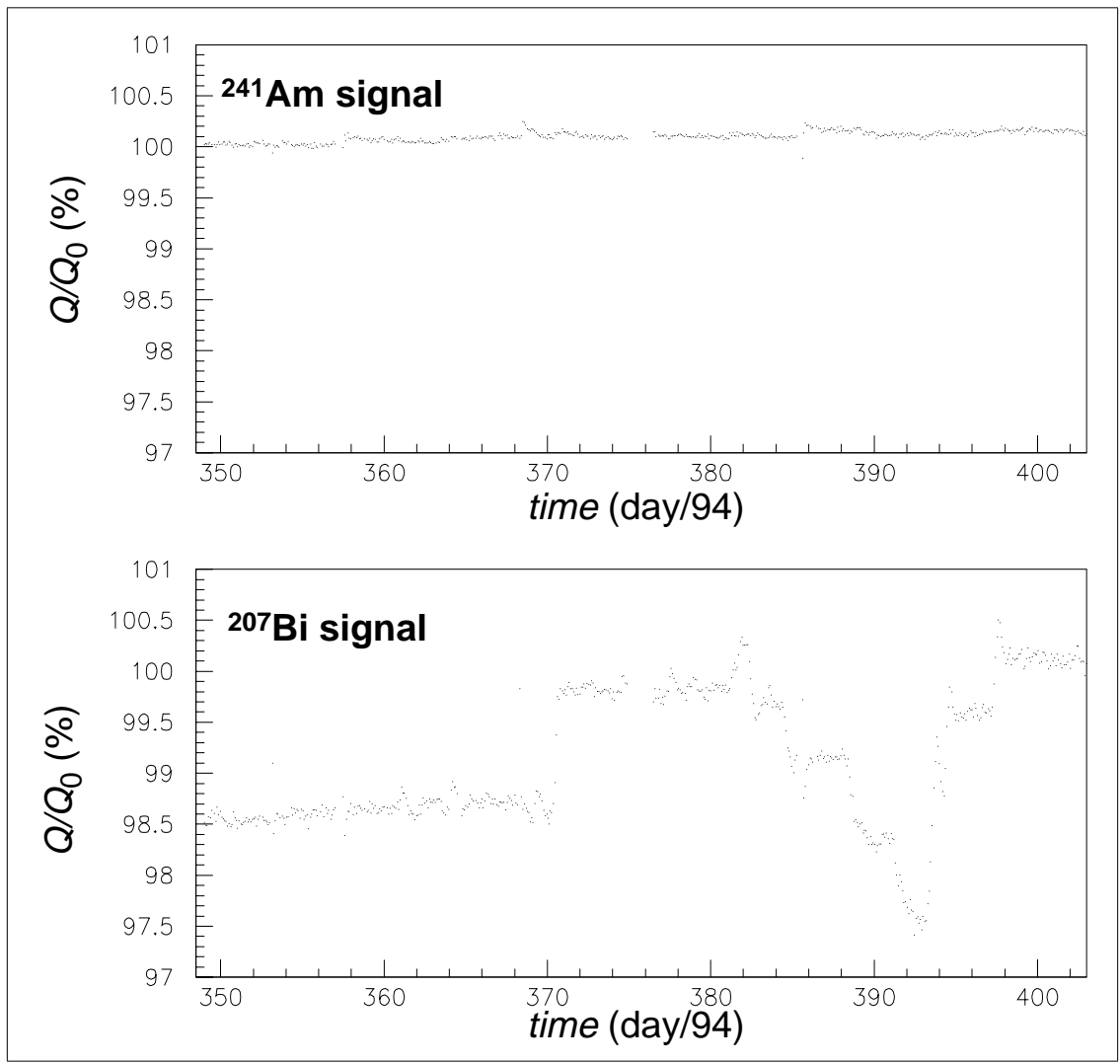

**Figure 6:** Comparison of $^{241}$Am and $^{207}$Bi charge collection efficiency $Q/Q_0$ during LAr-T scan

(equal to 0.035%).

The temperature gradient in the cryostat is not well known. Most likely systematic effects affecting the purity measurements at different points (or at different times during transients) can reach the 2 % level.

### 2.4 High voltage dependence of LAr-T effect

The analysis of high voltage dependence will be done in Section 3. Among our archive of high voltage curves, two are taken at the same purity level, with a "large" difference of temperature (1 °K). These 2 curves just differ by a change of yield (1.5%). We took this remark as a rule for our analysis of high voltage curves. The success of the temperature fit used in the cross-section scaling procedure (cf figure 12), proves that this rule represents also the effect of the smaller variations of temperature happening during the operation of H1.

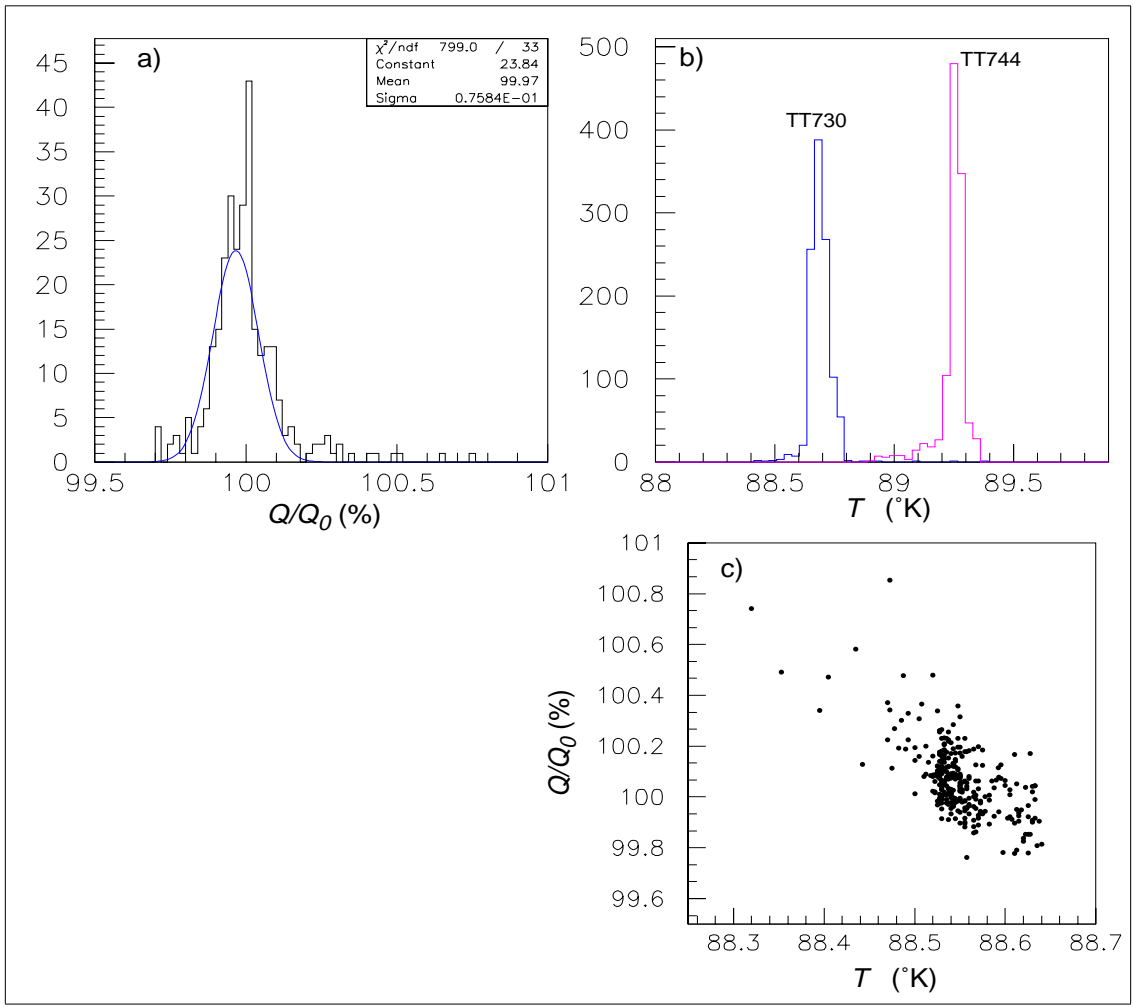

**Figure 7:** October 99: a) thermal fluctuation of the charge collection efficiency $Q/Q_0$ (probe 2; 1 MeV estimator); b) fluctuation of temperature measured in 2 points of the cryostat (TT730 and 744); c) charge collection efficiency of probe 2 vs. temperature (average of 6 points of the cryostat)

### 2.5 Discussion of the liquid argon temperature effect

The LAr-T effect, about 1.5%/°K, cannot be due to an artifact of the readout chain. Comparing alpha and beta sources gives us a strong constraint. It eliminates a direct effect of mobility, attachment or recombination of electrons. There are not many explanations left. If we suspect a "practical range" effect, we would expect it to be much smaller on a 482 KeV than on a 1 MeV line contrary to the facts. Moreover the range dependence at 1 MeV should be governed by the density variation (0.6% per °K) reduced by charge collection factors and even suppressed in principle for the 1 MeV estimator (cf.[4.0]). Another indirect effect of range is the modification of electric field due to the space charge of slowly drifting $Ar^+$ ions. The corresponding voltage drop, taking an ion/electron drift speed ratio $\sim 10^{-5}$ and energy deposition ~1 GeV/s, would be a few volts with a minimal effect on a beta source at 4 kV. It is interesting to note the temperature dependence of electron mobilities[3] in the high field range. A recent measurement fixes it at -1.7%/°K[12]. This is a new puzzle according to the recombination models presented in [6.1]: when temperature drops mobility increases. This should enhance the yield of alpha sources (not beta's) contrary to what we see. In summary, in order to explain the absence of LAr-T effect with alphas, recombination models should be revisited.

---

[3] The general trend described in [8] and [9] is complex at low field. The temperature coefficient changes sign at 0.1 kV/cm and has a peak at 140 °K

We cannot say presently if the LAr-T effect is a specificity of our purity probes or if it is generalizable to the H1 calorimeter, at least from test beam experience. More information from other calorimeter tests with high energy beams is needed.

## 3 High Voltage Curves

A high voltage curve depends on 2 important parameters in any simple model and for any type detector (cf.Appendix 2):

- an asymptotic charge $Q_0 = \lim_{E \to \infty} Q(E)$ related to the energy deposited in the detector
- a coefficient $E_0$, telling how fast $Q(E)$ tends towards $Q_0$

But it would be naive to think that these parameters have an absolute meaning. Well controlled experiments (cf.[9]) find significantly different values of $Q_0$ and $E_0$ according to the range of field strength retained for the fit. Some authors[4][11], following Thomas and Imel, look for a better fit by introducing a "*box model*"[3] supposed to represent recombination of electrons and ions and mixing it with Hofmann's model[7]. As it was pointed out by our late H1 colleague W.Flauger, they are doing it in vain because the shape of HV-curves is not altered significantly by adding recombination to attachment. This is demonstrated in Appendix 2, where we refute earlier "proofs" of the box model.

In this section we develop an analysis yielding consistent values of $Q_0$ and $E_0$ over the whole range of field strength and a calibration of the detector better than 0.2% RMS. It is based on a sequence of high voltage curves, taken periodically between 1991 and 1999 on two $^{207}$Bi probes (with respectively 4 and 6 mm gaps). During this period impurity concentration in the H1 cryostat has risen by a factor 5.

### 3.1 Scale invariance of impurity cross-sections

Purity and temperature are stable during the taking of a high voltage curve, because of the large volume of H1 liquid argon (56 m³). As seen in section [2], the charge yield $Q_0(T)$ changes by 1.4-1.8% per $^0$K. The impurity concentration $N$ has varied slowly during 8 years of operation. This gives us the opportunity to measure in 4 steps the real electron-impurity cross-section σ as a function of field strength $E$ defined by:

$$1/\lambda_e(E) = N \sigma(E) : \qquad (1)$$

1. <u>probe calibration</u> $\to Q_0(T_0)$ at given temperature and field strength
2. <u>temperature fit</u> $\to T$ for each curve
3. <u>scale invariance check</u> $\to$ impurity level $N$ constant with field strength
4. <u>impurity scale</u> $\to$ common $T$ and $N$ for all probes and H1 calorimeter

We define the **inverse function** ζ relating the 1 MeV charge estimator $Q$ and $\lambda_e$ by

$$Q/Q_0 = \lambda_e/D \times (1 - \exp{-D/\lambda_e}) \to D/\lambda_e = \zeta(Q/Q_0). \qquad (2)$$

where $\zeta(x) \approx 2\log 1/x \qquad x \to 1$ (3)

We have shown in paragraph 1.3 that at medium and high field strength Equation 2 gives $Q/Q_0$ within 1%. In order to measure the effect of a variable field strength on charge collection efficiency[4], we have changed the integration time from 2 μs to 3 μs. As it appears in figure 8, we consider that a 3 μs shaping time is safe for $E > 1$ kV/cm.

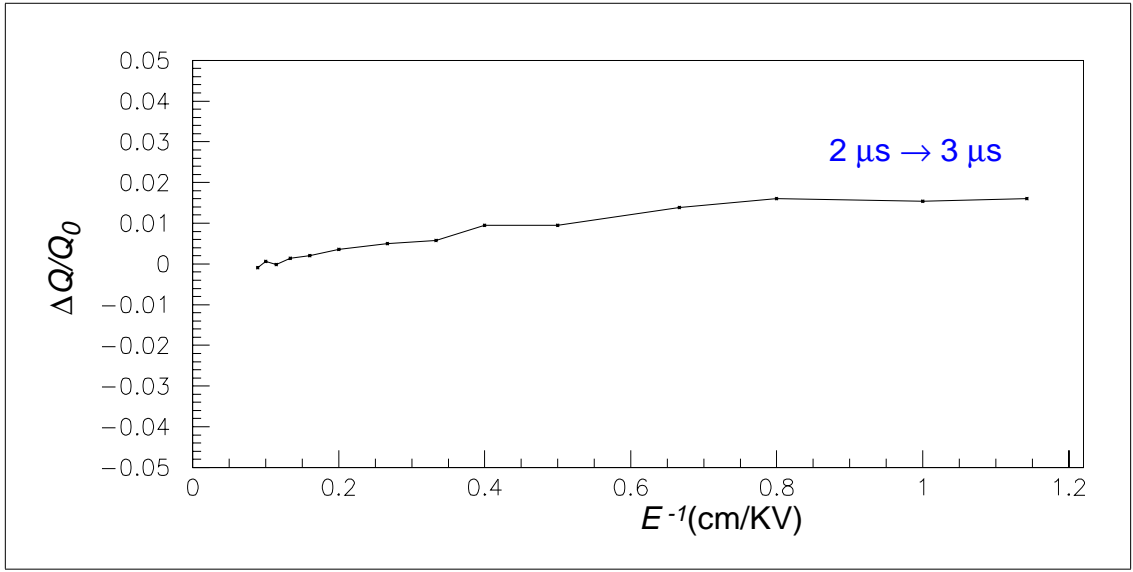

**Figure 8:** Increase of measured charge $Q$ (1 MeV) with shaping time versus field strength

The ζ function allows us to draw HV-curves in ($E^{-1}$, $D/\lambda_e$) instead of ($E,Q$) plane. Since $D/\lambda_e = DN\sigma(E)$, all HV-curves are proportional to $\sigma(E)$ when $N$ varies. Moreover a change of the calibration constant $Q_0$ is a translation parallel to the y axis in the approximation of Equation 3. Therefore the comparison of $D/\lambda_e$ in four points gives a calibration constant for each probe independently (in practice we made a fit on all points). The method is sketched on figure 9

Calibrating a probe is equivalent to finding a translation $D/\lambda_e \to D/\lambda_e - \delta y$ which makes the HV-curves y-scale invariant. Practically we have to compare the impurity concentrations $N_h$ and $N_l$ at 2 values of the field $E_h$ and $E_l$. In order to minimize statistical and systematic errors we used an averaging method. It consists in applying the same calibration procedure on the averages <$D/\lambda_e(E)$> of 2 sets of HV-curves (91-93 and 93-98), using the formula $\langle \zeta(Q(E)/(Q_0 + \delta Q_0)) \rangle = D\langle N \rangle \sigma(E) + \langle \delta y \rangle$, i.e. the linearity of purity and temperature perturbations.

Each probe being calibrated independently, we have tested as seen in figure 10 that the **ratio** of $\langle DN_h \sigma(E) \rangle = \langle D/\lambda_e(E) \rangle$ for the 2 probes is **1.5** = 6mm/4mm. (When necessary we have interpolated linearly the 4 mm probe data at the $E$-field values of the 6 mm ones). We have used 93-98 averages because: 1) the impurity concentrations being higher, the signal to noise is better than for 91-93 period, 2) during the 93-98 period, we have a set of 20 HV-curves taken for both probes at the same time.

Considering their good agreement, both probes are used to **determine the average electron**

---

[4] At low field (<1kV/cm), the maximum drift time being of the order of the shaping time of the electronics (3μs), the conditions of a pure charge integration are not met any more. One can expect a charge readout inefficiency depending on the detailed shape of the ionization current.

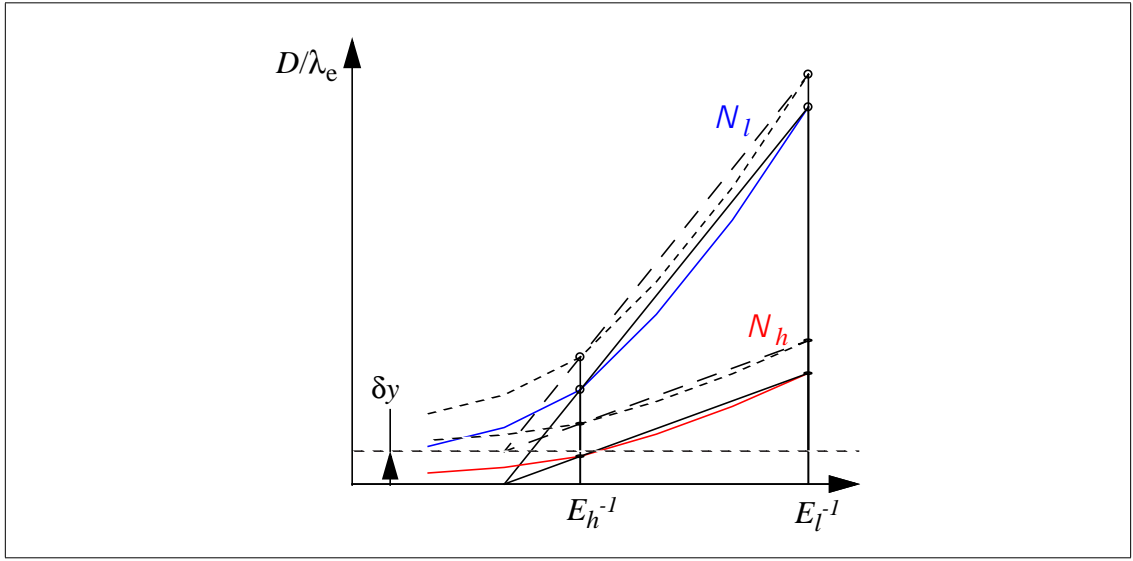

**Figure 9:** Properties of HV-curves in the ($E^{-1}$, $D/\lambda_e$) plane: 1) changing purity from $N_h$ to $N_l$ changes y-scale by a factor $N_l/N_h$; 2) changing calibration constant from $Q_0$ (plain) to $Q_0 + \Delta Q_0$ (dashed) translates curves by $\delta y$

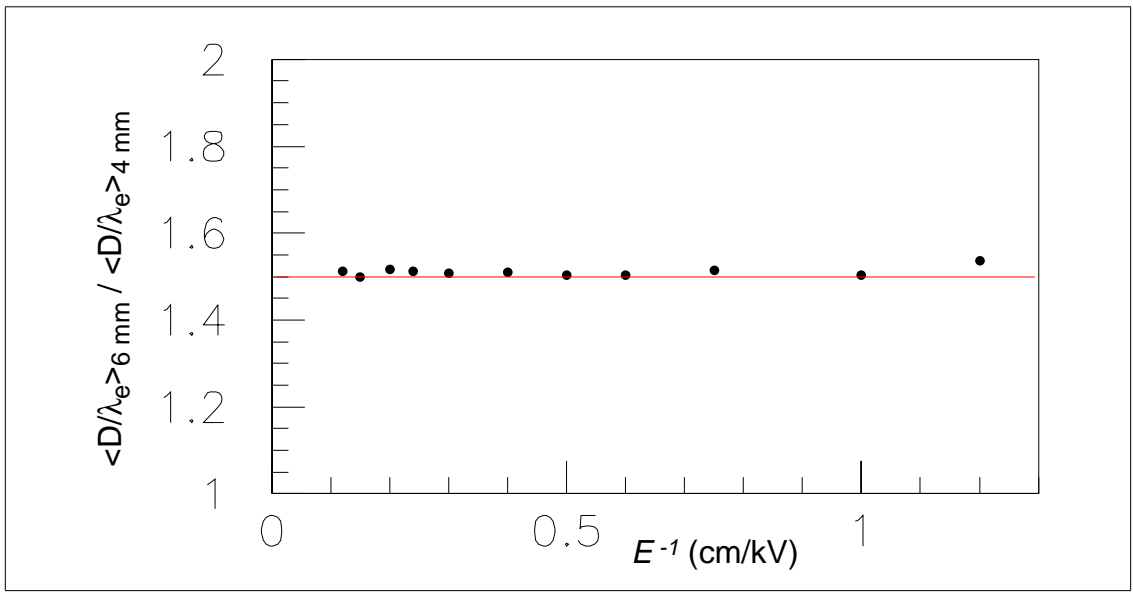

**Figure 10:** The ratio of relative absorption length in 6 and 4 mm probes compared to the gap ratio 1.5 ($<D/\lambda_e>$ averaged over 93-98 data). This is not a fit.

**mean free path** $\lambda_e(E)$ for the 93-98 period. In figure 11, 6 mm and 4mm data are mixed. When measured at the same field data points are superimposed perfectly, therefore one can fit a unique function $N_0\sigma_0(E)$ to these points. We have completed these data with 4 points between $E$= 10 and 15 kV/cm measured in year 2000. The remarkable linearity of $1/\lambda_e(E^{-1})$ on a large range of $E^{-1}$ is reproduced empirically by $\sigma_0(E^{-1}) = E_S E^{-1} + \exp{-E_a E^{-1}} - \eta$, which is linear for $E^{-1} > 2/E_s$ and is small for $E^{-1} \to 0$ due to the conditions $\eta \approx 1$ and $|E_a - E_S| \ll E_S$. Figure 11 shows that that fitting the $1/\lambda_e$ data with $N_0\sigma_0(E^{-1})$ gives an excellent result between $E$= 1 and 15 kV/cm, as well in the $1/\lambda_e$ vs. $E^{-1}$ representation (low field, fig.11a) as in $\lambda_e$ vs. $E$ (high field, fig.11b). Fitted values are $E_S$=12.0 kV/cm, $E_a$=15.7 kV/cm, $N_0$=0.929, $\eta$=1.004.

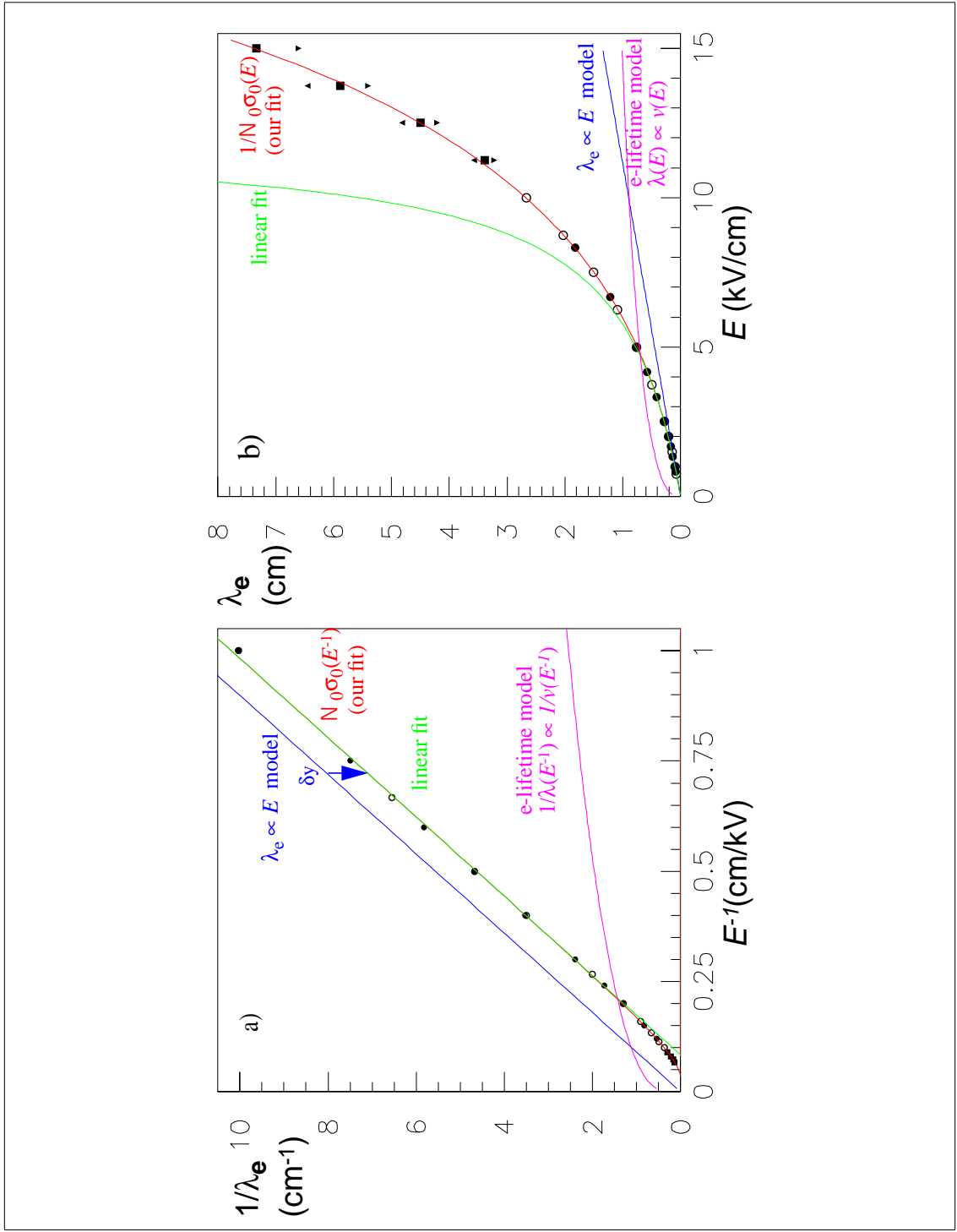

**Figure 11:** Four attachment models are compared to data in either $1/\lambda_e$ vs. $E^{-1}$ or $\lambda_e$ vs. $E$ plane. 1) our empirical formula, 2) $\lambda_e \propto E$ model from[7], 3) linear fit, 4) lifetime model (drift speed $v(E)$ from [12]). Data are $<1/\lambda_e>$ averages of 20 sets (year 93-98; probe 4 mm ○ or 6 mm ●) or extra high field points (year 00; 4 mm ■). LAr-T effect at high field +0.03⁰ (▲) or -0.03⁰ (▼)

Let us remark in the same figure that the classical *"electron lifetime model"* in which $\lambda_e(E) \propto v(E) \cdot \tau$ is incompatible with our data (lifetime $\tau$=constant; $v(E)$ taken from reference[12] is not linear in $E^{-1}$.

### 3.2 determination of an impurity scale

The empirical cross-section $\sigma_0 \propto 1/\lambda_e$ being arbitrarily normalized, we have defined two normalized cross-sections $\sigma_{O_2}$ and $\sigma_{H1}$ based respectively on:

1. $\lambda_e$ caused by one part per million of oxygen (ppm[$O_2$]) in LAr
2. $\lambda_e$ producing a 1% decrease of H1 electromagnetic calorimeter energy.

#### 3.2.1 oxygen impurity scale

The "*equivalent concentration of oxygen*" $N_{O_2}$ (ppm[$O_2$]), producing a given $\lambda_e$, is based conventionally on a $\lambda_e \propto E$ model parametrized by:

$$1/\lambda_{O_2}^{(cm)}(E) = N_{O_2}^{(ppm)} \sigma_{O_2}, \text{ with } \sigma_{O_2}(E)^{(cm^{-1} \cdot ppm^{-1})} = E^{-1}/0.15 \text{ }^5 \text{ for } E < 5 \text{ kV/cm.} \quad (4)$$

This model was introduced by Hofmann et al.[7]. Looking at figure 11a, we see that this model cannot fit our data. It needs first a $\delta y$ shift, i.e. a recalibration with a wrong calibration constant $Q_0$. This procedure is equivalent to a simultaneous fit of $Q_0$ and $N_{O_2}\sigma_{O_2}$ linear in $E^{-1}$. It yields a good $\chi^2$ only for $E < 5$ kV/cm and a $Q_0$ which depends on $N$.

If we define instead $\sigma_{O_2}(E) = \sigma_0(E)/0.15 E_S$ and $N_{O_2} = 0.15 \cdot E_S N_0 \quad (5)$

then $N_{O_2}\sigma_{O_2}$ fits our data for $1 < E < 15$ kV/cm and $N_{O_2}$ is a true number of ppm[$O_2$] equivalent. (For example the 93-98 average oxygen concentration is $N_{O_2}$=1.67 *ppm[$O_2$]*). In summary Hofmann's fit yields a wrong value of $Q_0$ which depends on $E$ and it has no predictive power in the 6 to 20 kV/cm range of the H1 calorimeter and probes. It cannot be used for the absolute calibration of a probe or a calorimeter.

#### 3.2.2 impurity scale of H1 electromagnetic calorimeter

We define $N_{H1}$=1% $u_{H1}$ ($u_{H1}$ stands for "H1 purity unit") as the increase of impurity yielding a $\Delta Q/Q_0$=1% decrease of the charge collected in the H1 electromagnetic calorimeter. Taking the nominal H1 parameters and a $\lambda \to \infty$ approximation from Table 1, we can introduce a function $\sigma_{H1}$ by $N_{H1}\sigma_{H1}(E)=1/\lambda(E)=(3/D).\Delta Q/Q_0$ and normalize it using the definition of the H1 purity unit, giving:

$$\sigma_{H1}(6.38)(cm^{-1} \cdot u_{H1}^{-1}) = 3/0.235. \quad (6)$$

Impurity scales are related by $N_{O_2}\sigma_{O_2} = N_{H1}\sigma_{H1} = N_0\sigma_0 = 1/\lambda \quad (7)$

which yields 1% $u_{H1}$ = 0.24 ppm[$O_2$]. The H1 impurity scale: a) uses the real cross-section $\sigma_0(E)$, b) gives the result needed for H1 calibration, c) is valid at any field strength, d) does not give the wrong impression that $O_2$ is the only pollutant.

**Table 1:** Parameters of H1 liquid argon cells at DESY

| source | [241]Am probe | [207]Bi probe | [207]Bi probe | EM calo |
|---|---|---|---|---|
| gap $D$ (cm) | 0.2 | 0.4 | 0.6 | 0.235 |
| field $E$ (kV/cm) | 19.5 | 6.5 | 6.0 | 6.38 |

---

5   the coefficient 0.15 ppm.cm²/kV comes from [7]. In [4] it is found to be 0.088

**Table 1:** Parameters of H1 liquid argon cells at DESY

| source | $^{241}$Am probe | $^{207}$Bi probe | $^{207}$Bi probe | EM calo |
|---|---|---|---|---|
| $^a log(Q_0/Q)^{(\lambda \to \infty)} \approx$ | $D/2\lambda$ | $D/2\lambda$ | $D/2\lambda$ | $D/3\lambda$ |
| $\sigma(E)/\sigma(E_{calo})$ | 0.8 | .97 | 1.11 | 1 |
| $G$ (Steiner factor) | $1.0^{exp}$ | 2.48 | 4.25 | 1 |

a. basic formulas and 1st order $1/\lambda$ approximations are found in Appendix 1 § 5.1

### 3.2.3 application of the H1 impurity scale

Any point-like charge collection estimator $Q$ at a field $E$ can be converted into an

impurity concentration estimator $N^{(u_{H1})} = \zeta((1 + \kappa\Delta T) \cdot Q/Q_0)/D\sigma_{H1}(E)$ (8)

(including the LAr-T correction $\kappa\Delta T)$. Using the logarithmic approximation of $\zeta$ this formula become $N = (\log Q_0/Q + \kappa\Delta T)/G$ for $\lambda \to \infty$, where the "*Steiner factor*" $G$ is a sort of "purity probe gain" which reads: $G = 3/2 \cdot (D/0.235) \cdot \sigma(E)/\sigma(6.38)$.

A first application the formula consists of converting each HV-curve into an impurity concentration curve. The cross-section scaling law implies that $N$ does not vary with $E$. This is well verified in figure 12, although several effects could violate this scaling:

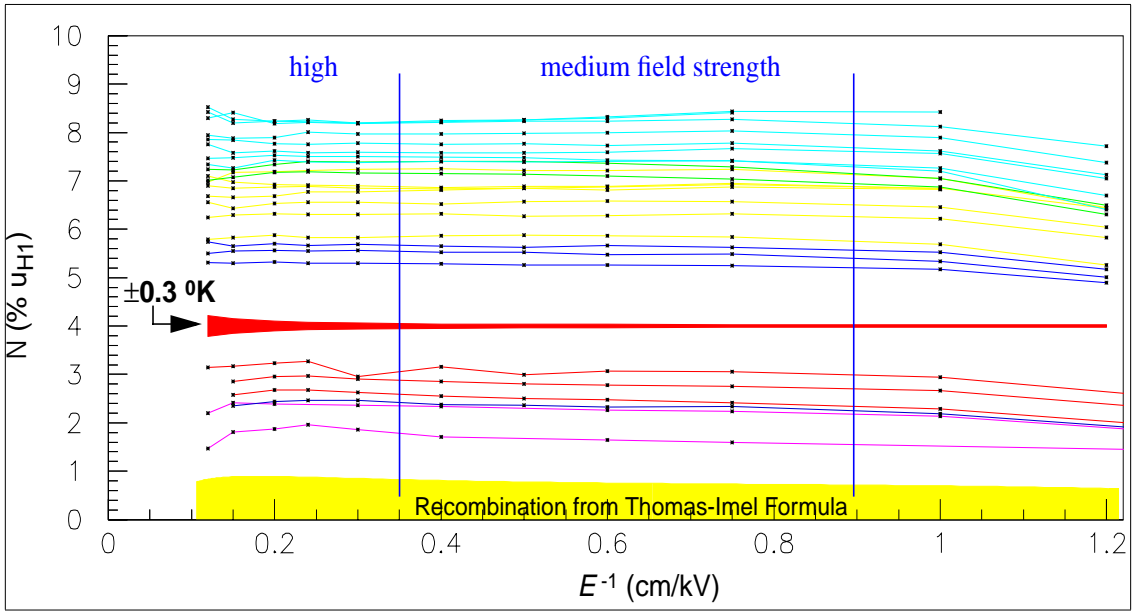

**Figure 12:** Impurity concentration curves for the 6 mm probe (1991-98). The medium field region defines an average impurity $N_m$ for monitoring H1 calorimeter over the years

- for E > 3 kV/cm the temperature correction is important
- for E < 1 kV/cm the absorption is too big, particularly for a 6 mm gap
- for $N_0$ < 2% the contribution of recombination is not well known
  (cf. fig.18: recombination for $^{113}$Sn [6] fitted by the box model of [3])

.Each impurity concentration curve can be resumed by a medium and high field average impurity, respectively $N_m$ for $0.35 < E^{-1} < 0.9$ cm/kV and $N_h$ for $E^{-1} < 0.35$. The dispersion of data around these average values is always < 0.1% $u_{H1}$ (RMS).
There is no difference between $N_m$ and $N_h$ basically because the exact temperature correction $\Delta T$ is fixed for each impurity curve by the condition $N_m - N_h = 0$. The effect of a $\pm 0.3\,°K$ shift around $N = 4\%\,u_{H1}$ is shown in figure . The difference between the $N_m$ measurements from the 4 and 6mm probes (shown in figure 13) is below $0.2\%u_{H1}$(RMS).

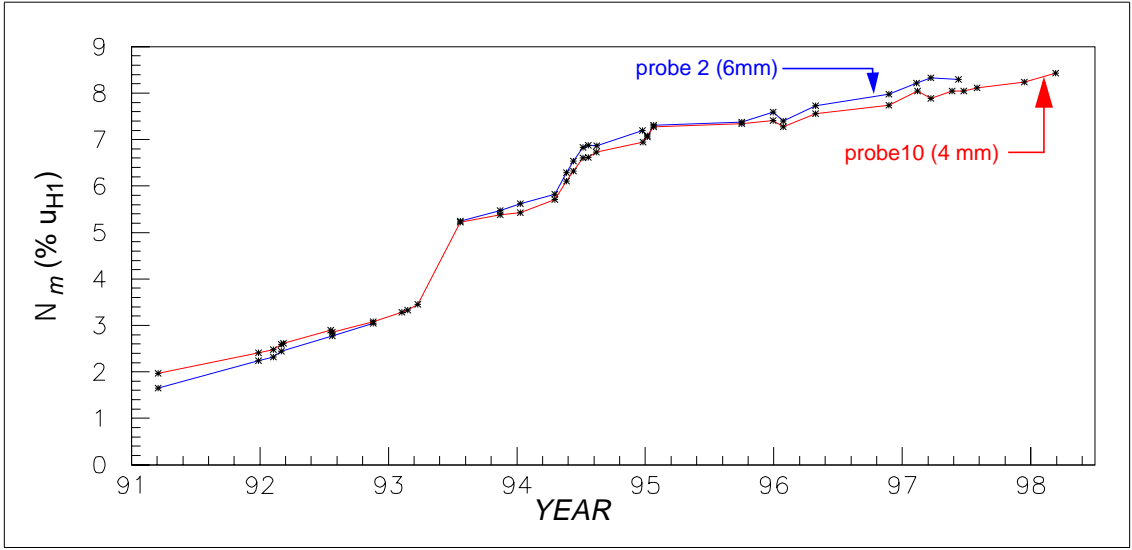

**Figure 13:** Impurity concentration increase in H1 from 91 to 98 seen by 4 and 6 mm probes.

A second application of impurity scales is the conversion of the impurity monitoring signals (1 MeV charge estimator in case of $^{207}$Bi) into impurity units as shown in figure 14. The approximation $N_{H1} = (\log Q_0/Q + \kappa\Delta T)/G$ is sufficient. However it requires the calibration constant $Q_0$, known for H1. It lacks an independent determination $\Delta T$ of the temperature correction although on a long period T can vary significantly. For example in figure 14, one has to know that the LAr temperature lost $0.3°K$ during the second half of year 94 in order to quantify the effect of the air leak.

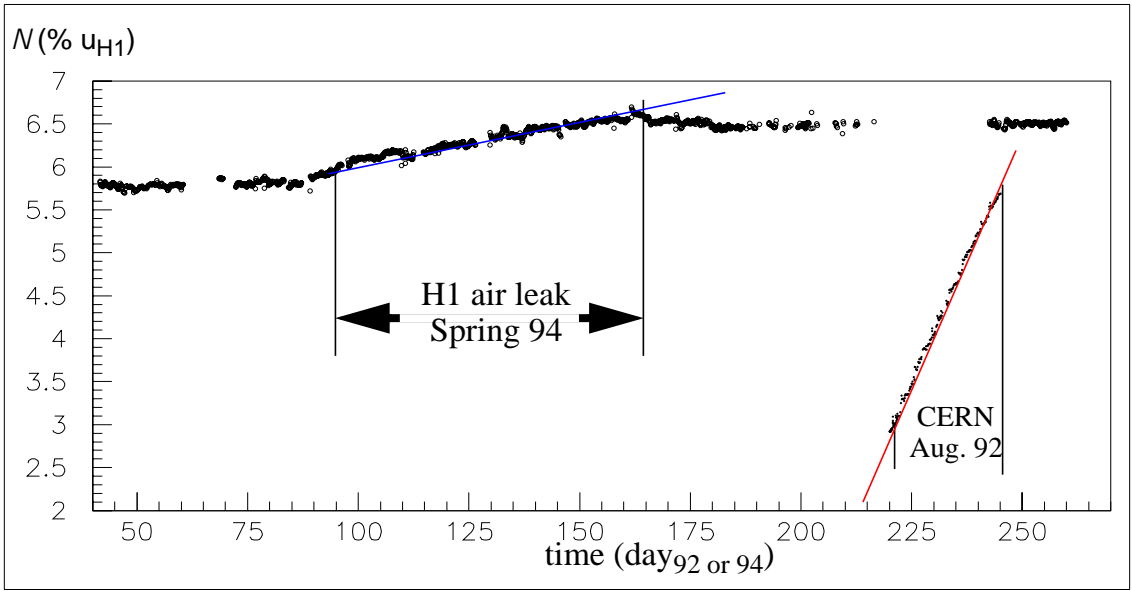

**Figure 14:** Impurity monitoring with $^{207}$Bi 6mm probes at $E \approx 6$kV (○ CERN 92; ● H1 94)

### 3.3 impurity scale at CERN beam tests

The purity monitoring system used on H1 was developed at CERN from August 89 to November 92, during tests of individual calorimeter modules. In 1992, during the last two running periods, it consisted of four [207]Bi and two [241]Am probes described in table [2] functioning like the H1 system installed at DESY in January 91. The fit of our empirical function $N_{H1}\sigma_{H1} = 1/\lambda(E)$ on the data of the HV-curve taken at CERN at the beginning of August 92 run is good, as seen in figure 15. It gives a calibration constant $Q_0$ which carried

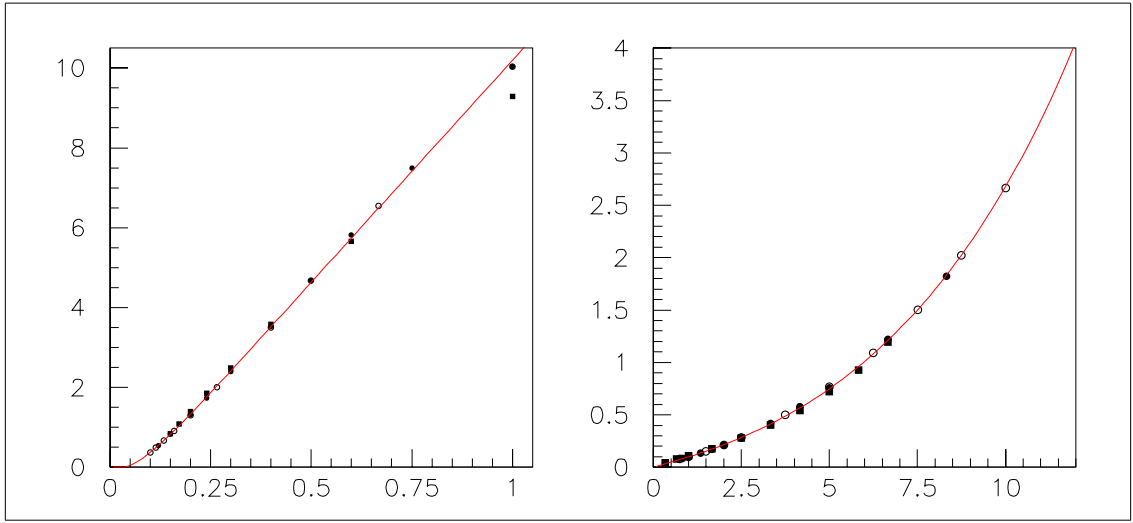

**Figure 15:** The function $N_{H1}\sigma_{H1}$ is fitted to CERN (Aug.92) and H1 (93-98 averages) data in either $1/\lambda_e$ vs. $E^{-1}$ or $\lambda_e$ vs. $E$ plane. 1) (CERN 6 mm ■ ; H1 4 mm ○ or 6 mm ●)

in the $N_{H1}$ formula of last paragraph transforms the charge $Q$ (1 MeV [207]Bi) recorded during August 92 run into the impurity monitoring data in figure 14. The regular increase of 3.5%$u_{H1}$/month was determined by the slope of a linear fit. By definition this factor converts the variation of the signal from a probe into H1 EM calorimeter variation. The variation of CERN EM calorimeter signal during August 92 was 1.8%/month. Therefore the measured ratio CERN/H1 is 1.8/3.5=0.52, to be compared with the predicted value $\sigma(10.6)/\sigma(6.38)= 0.58$. The slopes of the 5 other probes described in table 2 have been fitted similarly.

**Table 2:** Parameters of H1 liquid argon cells at CERN (August 92)

| source | [241]Am probe | [207]Bi probes | | | | EM calo |
|---|---|---|---|---|---|---|
| gap $D$ (cm) | 0.2 | 0.4 | 0.4 | 0.6 | 0.6 | 0.235 |
| field $E$ (kV/cm) | 20. | 12. | 10. | 8.33 | 6.66 | 10.64 |
| $log(Q_0/Q)^{(\lambda \to \infty)} \approx$ | $D/2\lambda$ | $D/2\lambda$ | | | | $D/3\lambda$ |
| $\sigma(E)/\sigma(6.38)$ | 0.63 | .50 | 0.58 | 0.78 | 0.96 | 0.58 |
| $G$ (Steiner factor)[a] | 0.8 | 1.3 | 1.5 | 2.9 | 3.7 | 0.52 |

a. $G$ is deduced from the slope of the signal decrease (for EM calo prediction is $G$=0.58)

They yield the experimental values of the "Steiner factor" $G$ and the corresponding values of $\sigma(E)/\sigma(6.38)$ found in table 2. (Let us remark that $\sigma(E)/\sigma(6.38)$ in tables 1 and 2 are larger for [241]Am at 20kV/cm than expected from the trend of fig.11b). Earlier the CERN system contained only two [207]Bi probes and its purity data were less precise and reliable. During a test period of 2 to 6 weeks, the calorimeter and the probes signal decreased due to a continuous release of impurities in liquid argon. Reversely the gain factors have been applied to the slopes measured during each CERN run. They yield the rises of impurity concentration seen in figure 16.

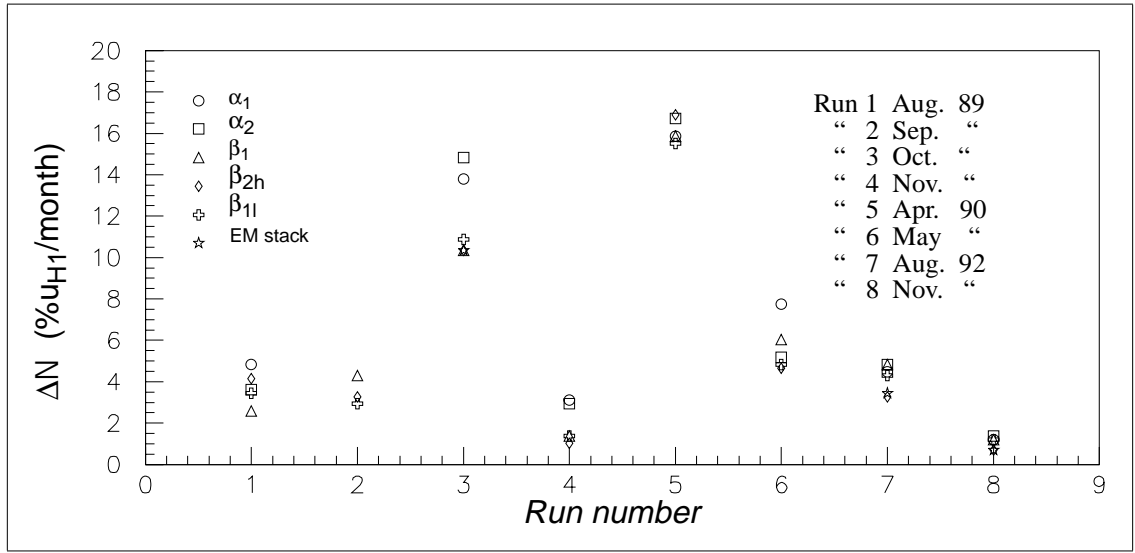

**Figure 16:** Rise of impurity concentration during CERN tests (% $u_{H1}$ / month) vs. run number

## 4  General conclusion

For the understanding of H1 calorimeter our work brings two sorts of information. The positive one first is that the long term variation of the calorimeter response due to pollution is well under control. Nevertheless it has to be corrected in order to compare the result of various years of data taking. Optimism come from the fact that the instable response of the various probes operating at the calorimeter standard field strength is well understood as being a result of **thermal fluctuations**. It can be precisely related to the response at other field strength owing to the determination of the real **electron attachment cross-section** instead of approximate models. Measurement done by two different probes are consistent within ≈0.2% RMS.

The negative information is that we do not know the exact sensitivity of H1 electromagnetic calorimeter to temperature which could be similar to our probes. In this case the spatial fluctuations of the calorimetric response due to temperature gradients could be in the 2% range. As long as experiment has not concluded there will be some doubt, considering the incompatibility of our liquid argon temperature effects with the prevailing recombination models.

At last it seems to us that the facts are solid and strange enough to provoke some experiences aiming to understand better the ionization processes in liquid argon.

## Acknowledgments

To make a long story short, H1 liquid argon purity system is the child of T.P.Yiou. Probes were constructed and installed in a cryostat, first in Paris, secondly at CERN and finally in HERA north hall by LPNHE-Paris technical team, under D.Imbault leadership, with diligent participation of B.Canton, J.David and P.Repain. The prototype analog and readout chain was set up and tested at CERN by F.Descamps and later built and integrated in H1 environment by P.Bailly, J.F.Genat and F.Rossel for the electronics and F.Martin for the readout. H.Steiner came at the right time and did the right things to help understanding and controlling the "H1 purity crisis". He initiated with W.Flauger and K.Thiele the portable purity system (not covered by this paper). Then the supervision of the purity system, vital for H1 calorimeter, was taken over progressively by G.Villet and A.Babaev. The easy web access to the purity and temperature data is due to A. Cyz, M.Hensel and J. Martyniak. Collaboration with H1 Cryostat team was essential for our analysis.

This survey gives me the occasion to express my gratitude to H.Oberlack, G.Cozzika and M.Fleischer for their past support, and to offer my best wishes to J.Ferencei who now bears the brunt.

# 5 Appendix 1

## Charge collection efficiency

### 5.1 single track charge collection efficiency

5.1.1 Assumptions:

We are using a classical charge integration electronic chain. Our spectral analysis is done at high field (>>1KV/cm). The integration time (3μs) of the electronics is longer than the electron drift from cathode to anode (distance $D$). Ions are not supposed to move.

5.1.2 Computation of the charge collection ratio $Q/Q_0$:

The <u>relative position</u> $\xi$ in a gap is defined by $\xi = x/D$, where the $x$ axis is perpendicular to the cathode; $\xi=0$ on the cathode and $\xi=1$ on the anode; $0 \le \xi \le 1$ in general.
The <u>drift of an electron</u> at instant $t$ is $\xi(t) = \xi_0 + v_d t/D$ ($v_d$ is the drift speed)

The expression $\dfrac{d^2Q}{d\xi_0 dt} = q(\xi_0) \cdot A(t)$ represents the electric flow produced at instant t by the drift of the ionization density $dQ/d\xi_0 = q(\xi_0)$ deposited at $t=0$ around $\xi_0$.

The disparition of charges, either vanishing on the anode (step function $\theta$) or attached to impurities (exponential attenuation length $\lambda_e$), is modeled by an attenuation function
$A(t) \to e^{-(\xi - \xi_0)(D/\lambda_e)} \cdot \theta(1 - \xi)$ after the change of variable $t \to \xi$.

The total deposited charge is $Q_0 = \int_0^1 q(\xi_0) d\xi_0$ and the collected charge is

$$Q = \int_0^1 d\xi_0 \int_{\xi_0}^1 \dfrac{d^2Q}{d\xi_0 d\xi} d\xi = \int_0^1 q(\xi_0) d\xi_0 \int_{\xi_0}^1 e^{-(\xi - \xi_0)(D/\lambda_e)} d\xi \qquad (9)$$

or after one integration

$$Q = \dfrac{\lambda_e}{D} \int_0^1 q(\xi_0)(1 - e^{-(1 - \xi_0)D/\lambda_e}) d\xi_0 \qquad (10)$$

The "measured charge" $Q$ depends only on the drift distance, not on the drift speed $v_d$.

5.1.3 Integration of $Q/Q_0$ for different track geometries.

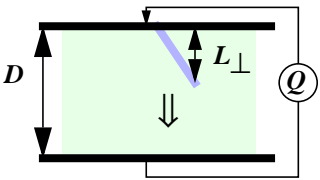

- **track origin on the cathode & uniform ionisation density**:
Integrating previous formula with $q(\xi_0) = Q_0 D/L_\perp$ and $0 < \xi_0 < L_\perp/D$ gives:

$$Q/Q_0 = \eta\left(\dfrac{L_\perp}{D}, \dfrac{\lambda_e}{D}\right) \qquad L_\perp \le D$$

with an attenuation function defined as[6]:

---

[6] at singular points $\eta$ is defined by continuity, e.g. $\eta(0,v) = \lim \eta(u,v)$ when $u \to 0$

$$\eta(u, v) = v\left[1 - \frac{v}{u}\left(1 - e^{-\frac{u}{v}}\right)e^{-\frac{1-u}{v}}\right] \tag{11}$$

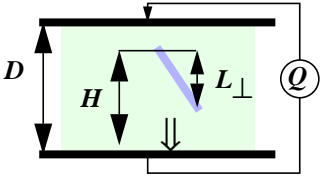

- **track origin at a distance $H$ from the anode**:
Integrating with $q(\xi_0)=Q_0 D/L_\perp$ and $1-H/D<\xi_0<1-H/D+L_\perp/D$ gives:

$$Q/Q_0 = \frac{H}{D} \times \eta\left(\frac{L_\perp}{H}, \frac{\lambda_e}{H}\right)$$

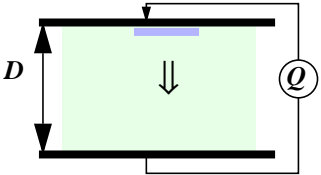

- **point source on the cathode**

$$Q/Q_0 = \eta\left(0, \frac{\lambda_e}{D}\right) = \frac{\lambda_e}{D} \times (1 - e^{-D/\lambda_e}) = e^{-D/2\lambda_e} \times \frac{\sinh D/2\lambda_e}{D/2\lambda_e}$$

behaviour at high purity: $Q/Q_0 \sim 1 - \frac{D}{2\lambda_e}$ for $\lambda_e/D \to \infty$

id. at 2nd order: $Q/Q_0 \sim e^{-D/2\lambda_e} \to D/\lambda_e = \zeta(Q/Q_0) \sim 2 \times \log(Q_0/Q)$

behaviour at low purity: $Q/Q_0 \sim \frac{\lambda_e}{D}$ for $\lambda_e/D \to 0$

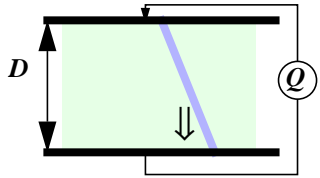

- **track crossing the whole gap**:

$$Q/Q_0 = \eta\left(1, \frac{\lambda_e}{D}\right) = \frac{\lambda_e}{D} \times \left(1 - \frac{\lambda_e}{D}(1 - e^{-D/\lambda_e})\right)$$

behaviour at high purity: $Q/Q_0 \sim \frac{1}{2} \times \left(1 - \frac{D}{3\lambda_e}\right)$ for $\lambda_e/D \to \infty$

behaviour at low purity: $Q/Q_0 \sim \frac{\lambda_e}{D}$ for $\lambda_e/D \to 0$

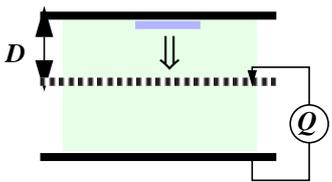

- **point charge on the cathode reaching a grid at intermediate voltage** (Biller et al.[4]):

$$Q/Q_0 = e^{-D/\lambda_e} \quad \text{(D varies, Q measured from grid to anode)}$$

behaviour at high purity: $Q/Q_0 \sim 1 - \frac{D}{\lambda_e}$ for $\lambda_e/D \to \infty$

behaviour at low purity:

$$Q/Q_0 \ll \left(\frac{\lambda_e}{D}\right)^n \quad for \quad \lambda_e/D \to 0$$

$$Q/Q_0 \cong 0.53\left(\frac{\lambda_e}{D} - 0.242\right) \quad for \quad 0.3 < \frac{\lambda_e}{D} < 0.6$$

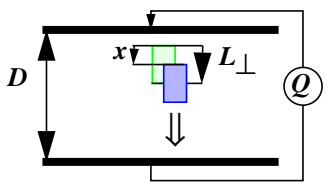

- **box model**:
initially $Ar^+$ and $e^-$ boxes are superimposed ($x=0$) and the $e^- + Ar^+ \to Ar$ mean free path is $\lambda_0$. As long as $x<L_\perp$ boxes overlap. Charge densities $n^+(x)=n^-(x)$, uniform inside $[x,L_\perp]$, are normalized by $n^\pm(0)L_\perp=Q_0$. They verify the differential equation:

$$dn^- = -n^+ n^- \frac{dx}{\lambda_0} \Rightarrow d\left(\frac{1}{n^-}\right) = \frac{dx}{\lambda_0} \Rightarrow \frac{1}{n^-(x)} - \frac{1}{n^-(0)} = \frac{x}{\lambda_0} \Rightarrow n^-(x) = \frac{n^-(0)}{1 + x/\lambda_0} \quad (12)$$

Out of the ion box ($x > L_\perp$), e$^-$ do not interact any more. The final charge is:

$$Q = \int_0^{L_\perp} n^-(x) dx \Rightarrow \frac{Q}{Q_0} = \frac{\lambda_0}{L_\perp} \log\left(1 + \frac{L_\perp}{\lambda_0}\right) \quad (13)$$

### 5.2 $^{207}$Bi spectral function

We reproduced the $^{207}$Bi energy spectrum using a spectral function varying with gap size and impurity concentration. This function $S_{Bi}(q)$ represents the number of events yielding a charge $Q$ in a [**q**, **q**+δ**q**] interval during a time **ΔT** (δq= 1 u$_{ADC}$; ΔT= 10'):

$$S_{Bi}(q) = \int_0^1 d\xi \sum_{1,2,3} \alpha_i G(q_i^C(\xi)/\kappa_i \sigma_i) + [\beta_i G(q_i^K(\xi)/\sigma_i) + \delta_i G(q_i^L(\xi)/\sigma_i)] w(\xi) \quad (14)$$

$$G(t) = e^{-t^2/2}/\sqrt{2\pi}$$

**α$_i$**, **β$_i$** and **δ$_i$** are the number of gammas leaving an energy $E_{\gamma_i}$, respectively by Compton scattering, by conversion of electrons of the K and of the L layer, during ΔT.
The RMS resolution **σ$_i$**, affecting the measurement of the charges $q_i^{K,L}(\xi)$ deposited by K and L electrons, is supposed to be Gaussian. It is a quadratic combination of the electronic noise measured on the calibration peak σ$_0$ =12 u$_{ADC}$ ↔ 400e$^-$ and of a "constant term" σ$_E$ /E= 3.2%. The RMS resolution on the Compton energies $q_i^C(\xi)$ is made a factor κ$_i$ larger than noise (κ$_1$=κ$_2$=1.4; κ$_3$=5.3) in order to incorporate the finite width of the Compton edge.

Within our model, the charges $q_i^C$ and $q_i^{K,L}$ are related to energies by:

$$q_i^C = q - \left[q_0 + s \cdot E_i^{edge} \cdot k \cdot \eta(0, \frac{\lambda_e}{kD})\right] \quad (15)$$

with $E_i^{edge} = 2E_{\gamma_i}^2/(m_e + 2E_{\gamma_i})$ \quad (16)

$$q_i^{K,L} = q - \left[q_0 + s \cdot (E_{\gamma_i} - E_{K,L}) \cdot \eta(\frac{h \sin\theta}{D}, \frac{\lambda_e}{D})\right] \quad (17)$$

with $\eta(u, v)$ defined in Equation 11.

In these formulas **q$_0$** is the ADC offset, **s** the ADC to energy ratio, **E$_K$** and **E$_L$** (88 and 13 KeV respectively) are the binding energies of K and L electrons in bismuth, $\eta(h\sin\theta/D, \lambda_e/D)$ is the charge collection efficiency for an uniform electron track emitted on the cathode, at an angle θ, with a path **h**, in a gap of width **D** filled with LAr where the electron mean free path is $\lambda_e$, $E_i^{edge}$ the energy of the Compton edge of photons γ$_i$. We have assumed that Compton electrons are roughly parallel to the anode. Then the charge collection efficiency $\xi\eta(0, \lambda_e/\xi D)$ depends only on the distance ξa from the plate. In order to integrate the ξ variable literally in Equation 14 the Compton spectrum is approximated by a delta peak at the edge plus a step function below the edge. It yields:

$$S_i^{compton}(q) = \sum_{1,2,3} (A_i + B_i \cdot q_i^C(1)) freq((q_i^C(1)/\kappa_i \sigma_i)) \qquad (18)$$

$$freq(x) = \int_{-\infty}^{x} G(t)dt$$

The integration variable $\xi = h\sin\theta/R(E)$ for K an L electrons is the distance from the origin to the end-point of an electron track of energy $E$, divided by the so-called "practical range" $R(E)$. Classically the end-point is distributed uniformly <u>inside</u> (not on the surface) of a sphere of radius $R(E)$. This is represented by a repartition function $W(\xi)$:

$$dW(\xi) = w(\xi)d\xi = d\left(\frac{3\xi - \xi^3}{2}\right) \text{ or almost equivalently by } dW(\xi) \approx d\left(\sin\frac{\pi}{2}\xi\right) \qquad (19)$$

# 6 Appendix 2

## attachment and recombination models

### 6.1 Processes causing electron losses

The charge collected $Q(E)$, for a drift field $E$, is derived from the deposited charge $Q_0$ after taking into account 3 kind of electron losses.

#### 6.1.1 recombination of geminate pairs

The initial electron recombination in a *"geminate pair"* (electron-ion pair) depends on the electron thermalization length and the angle made by its momentum and the drift field. The average electron escape probability is given as a function of $E$ in the Onsager model[1] by:
$Q/Q_0 \cong \alpha(1 + E/E_{Ons}) \qquad for \qquad E \to 0$
where $\alpha$ and $E_{Ons}$ are related to thermodynamical variables.

#### 6.1.2 recombination inside an ion cloud.

The electron, after escaping from its parent ion, is recombined to an other ion of the same track. There are 2 models of *"ion clouds"* (or *"blobs"*):

<u>Jaffé's "columnar" model</u> [2]:
general formula:   $Q/Q_0 = 1/(1 + E_{Col}/E)$ (20)

behaviour at high field: $Q/Q_0 \sim (1 - E_{Col}/E) \qquad for \qquad E \to \infty$ (21)

behaviour at low field:   $Q/Q_0 \sim E/E_{Col} \qquad for \qquad E \to 0$ (22)

<u>Thomas-Imel "box" model"</u>[3]:
general formula:   $Q/Q_0 = E/2E_{Box}\ln(1 + 2E_{Box}/E)$ (23)

behaviour at high field is: $Q/Q_0 \sim 1 - E_{Box}/E \qquad for \qquad E \to \infty$ (24)

behaviour at low field:

$$Q/Q_0 \sim E_1/2E_{Box} + \delta E \ln(1 + 2E_{Box}/E_1) \qquad for \qquad \begin{pmatrix} E_1 \to 0 \\ E = E_1 + \delta E \end{pmatrix} \qquad (25)$$

This model is based on a distribution of electrons and ions uniform within a box (cf. Equation 13) and a recombination cross-section proportional to the field strength: $L_\perp/\lambda_0 = 2E_{box}/E$.

Discussion of recombination models :

For a given α or β source, the parameters $E_{box}$ and $E_{col}$ should be identical. Coming both from a [113]Sn (380 keV) source, the parameters $E_{box}$= ξ$E/2$= 0.42 kV/cm found in reference[3] and $E_{col}$= 0.53 kV/cm found in reference[6] should be the same within errors.
On the contrary these recombination parameters vary wildly for different sources and different liquids. For instance in liquid xenon Thomas[3] found $E_{rec}$= 0.075 kV/cm for [113]Sn (380 keV) and Séguinot[10] $E_{rec}$= 1.83 kV/cm for an electron beam (2-40keV). In liquid argon Thomas[3] found $E_{rec}$= 0.42 kV/cm for [113]Sn and $E_{rec}$= 280 kV/cm for α particles. In conclusion:

- The recombination models should be considered only as a rough approximation, valid when the probability of recombination is small. They could even be incompatible with the LAr-temperature effect as reported earlier in Section 2.5.
- The hypothesis that recombination cross-sections are proportional to 1/E is not supported experimentally
- One should not use the constant ξE=0.84 kV/cm of Thomas[3], fitted on [113]Sn data, for a [207]Bi source, as done in[4]. (We have not found a determination of this [207]Bi constant in the literature).

### 6.1.3 attachment on electro-negative impurities

1. The *real attachment cross-section* σ($E$) has been determined in Section 3.1 with a concentration $N$ of impurity molecules as: $D/\lambda_e = DN\sigma(E)$
2. The *electron lifetime model* considers that the probability of electron attachment along a path depends only on drift time. Let us define the drift field $E_D$ in which the attachment mean free path $\lambda_e$ is equal to the distance $D$ from anode to cathode and the mobility μ($E$). Then the number of electron mean free path in the gap $D$ is: $D/\lambda_e = (E_D/E) \times \mu(E_D)/\mu(E)$.
3. The *constant mobility model* since Hofmann[7] consider mobility μ($E$) as a constant (in reality, cf. Miller[5][12], mobility vary by a factor 2 from $E$=1 to 10 kV/cm). This is expressed by: $D/\lambda_e = E_D/E$
   Hofmann's formula[7] is obtained by carrying this into the "point source" charge collection of [5.1]. For $\lambda_e \to \infty$ its yields: $Q/Q_0 \sim 1 - E_{att}/E$, to be used in [6.2.1] with $E_{att}=E_D/2$

In figure 11 the real cross-section is compared with the lifetime and the constant mobility models, either in the σ($E^{-1}$) or the λ($E$) representation. It shows that the constant mobility is somewhat better justified than lifetime hypothesis, although the former was introduced as a consequence of the latter.

## 6.2 Recombination and attachment mixed

The proposition of Thomas and Imel[3] to multiply the point source attachment formula of Hofmann[7] by the box recombination formula has been widely followed [4][11]. More generally it is reasonable to **multiply the 3 charge attenuation factors** corresponding to

1. geminate pair recombination,
2. ion cloud recombination,
3. attachment on impurity,

because these 3 processes are independent.
We shall now prove that, supposing this hypothesis true, it is almost impossible to unfold the 3 contributions only by fitting the shape of $Q(E)$.

### 6.2.1 High drift field

In this range the box recombination and the attachment models give charge attenuation factors respectively of the following form: $Q^{mix}/Q_0 \sim 1 - (E_{box} + E_{att})/E$, which combine into:

$$Q^{mix}/Q_0 \sim 1 - (E_{box} + E_{att})/E \tag{26}$$

This means the same $E^{-1}$ asymptotic branch for pure attachment, pure recombination or mixed attachment and recombination.

### 6.2.2 Medium drift field

We have to emphasize the fact that mixtures of box recombination and attachment with the same high drift field behavior (because of $E_{box} + E_{att} = E_{mix}$) are almost indistinguishable also in the medium field range, defined by $Q^{mix}/Q_0 > 0.2$. $\tag{27}$

This mathematical ambiguity is illustrated in figure 17b, where we vary the proportion of $E_{box}$ in the mixture from 0% to 80%. The drift field is varied in order to have a charge collection efficiency running from $Q/Q_0$=0 to 93% and the charge normalization error is kept under 2%. The discrepancy between pure box model recombination and no recombination at all is always below ±0.7% for $Q/Q_0$>0.45 and below 2% for $Q/Q_0$>0.2. Comparing this mathematical discrepancy function to the systematic errors in the best available data[6], shows that $\chi^2$ does not test the validity of the recombination model.

### 6.2.3 Low drift field

The drift fields yielding charge collection efficiencies $Q/Q_0 < 0.2$ are not relevant for our analysis. Firstly because the product of 3 processes (each linear in $E$) mentioned in [2.1] give a $E^3$ behavior which is not seen in the data except maybe in[9]. Secondly, one should develop experimental techniques working at low field as did Eibl et al[8] (long drift time, photo injection source) or Séguinot et al[9] (pulsed electron sources), just because radioactive electron sources have a bad signal over noise ratio in this range. For example a $^{113}$Sn source would yield ~2500 e-, at best 10 $\sigma_{noise}$.

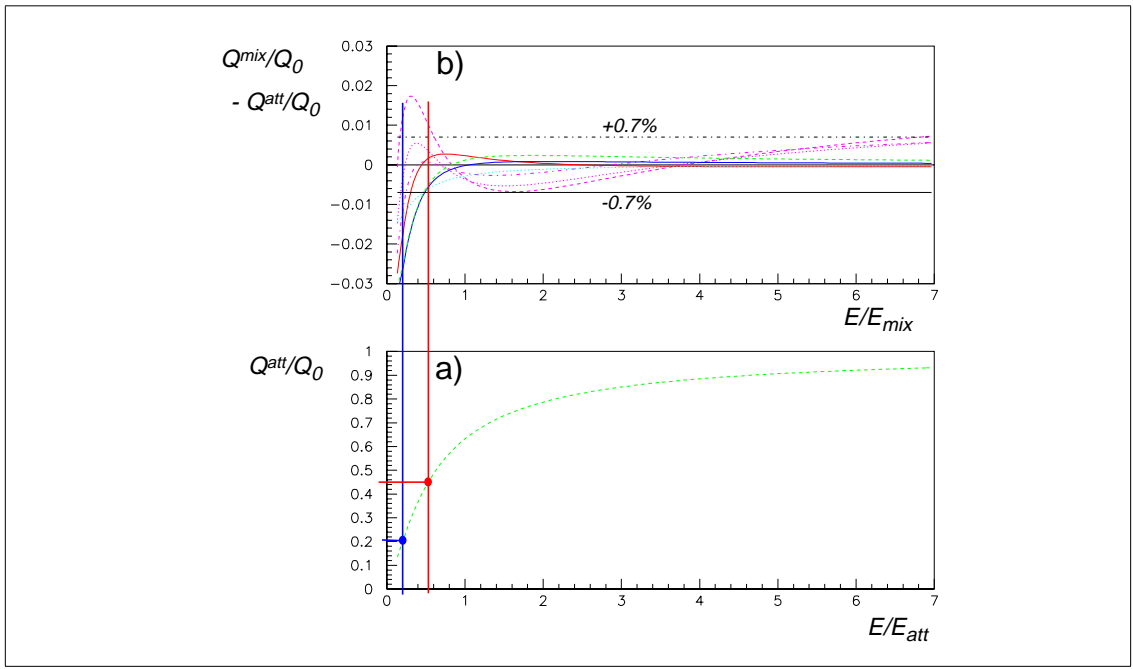

**Figure 17:**  Comparaison between simple Hofmann model and a composite one (Hofmann+box)
a) $Q^{att}/Q_0$ as a function of $E/E_{att}$ from Hofmann's formula;
b) $Q^{mix}/Q_0 - Q^{att}/Q_0$ as a function of $E/E_{mix}$ ($E_{att}=E_{mix}$)

### 6.2.4  experimental lack of evidence

The agreement of the box model with the data is not as good as often claimed. Superposing in figure 18 b Thomas' fit of Scalettar's[6] data and the original Scalettar's figure, we see discrepancies an order of magnitude larger than the <0.7% discrepancy between a box model and no box model (fig.17).

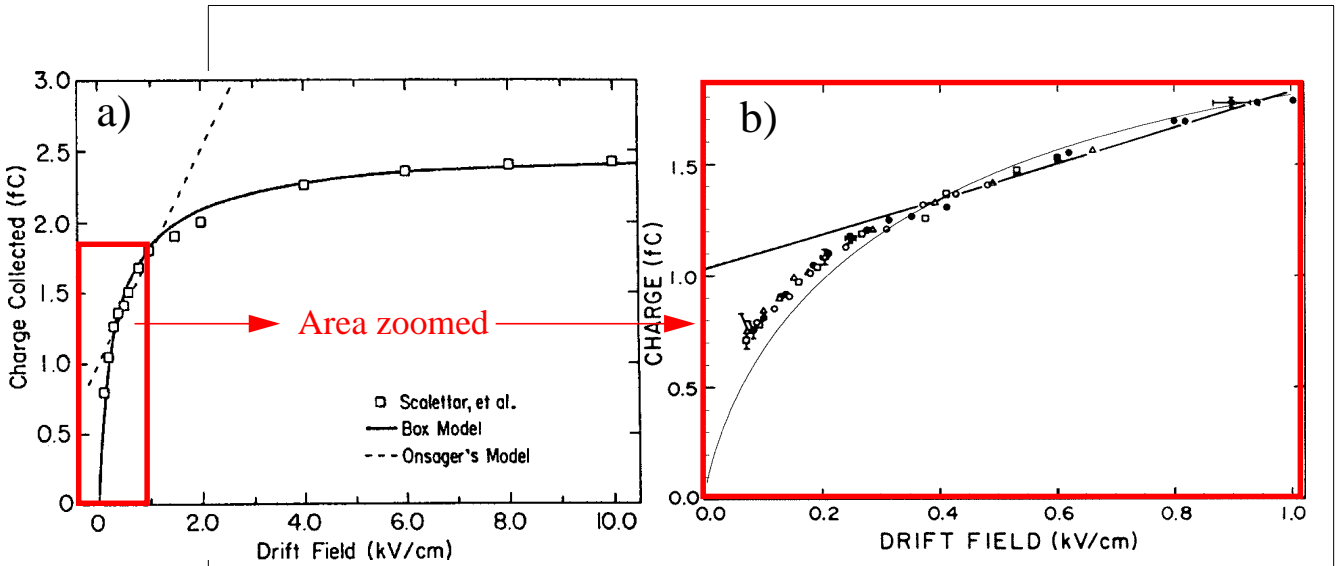

**Figure 18:**  a) Original figure of Thomas-Imel[3] paper showing the result of its box model fit
b) True comparison of Scalettar's[6] original data with "box model" fit of Thomas[3]

We reproduce in figure 19, the figure 7 of reference [4] which comments *"comparison with the data*

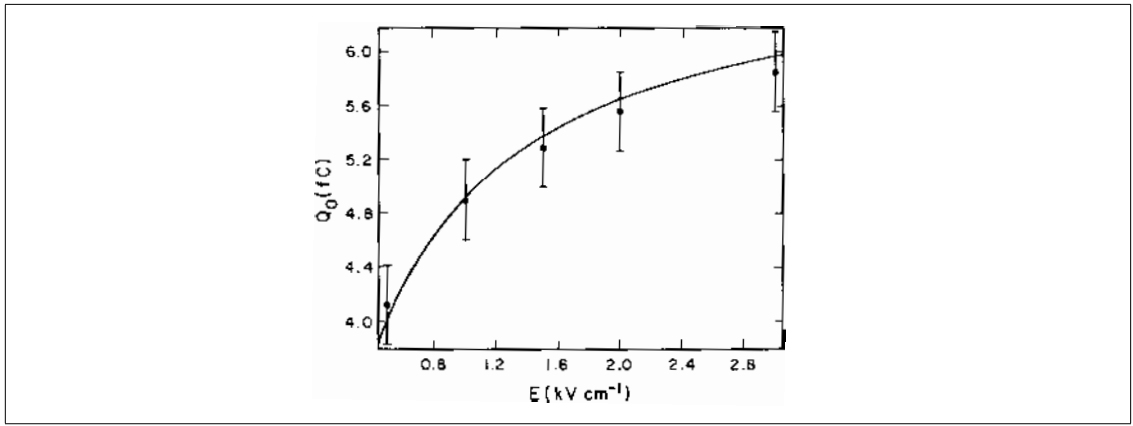

**Figure 19:**   Biller's box model test using $^{207}$Bi data. (figure copied from reference[4])

*yields a reduced $\chi^2$..."*. This fit is the only support of the use of $^{117}$Sn box recombination parameter with $^{207}$Bi experimental data.

6.2.5   Conclusion on models mixing recombination and attachment

Recombination and attachment models yield very similar field dependence, due to their common assumption making cross-sections inversely proportional to the electrical field. The characteristic $E_{\text{mix}}$ parameter of a mixed model is obtained by adding those of its recombination ($E_{\text{box}}$) and attachment ($E_{\text{att}}$) components. The recombination parameters should be measured with different sources (radioactive or calorimetric). The asymptotic behavior of the recombination $E\rightarrow$ should not be taken for granted. More generally the effects described in section [2] seem in contradiction with the models presented above.